\documentclass[12pt]{article}

\usepackage[utf8]{inputenc}
\usepackage{graphicx}
\usepackage{enumitem}
\usepackage{amsmath, bm, amsthm, esint}
\usepackage{listings}
\usepackage{hyperref}
\usepackage{float}
\usepackage{comment}
\usepackage{xcolor}
\usepackage{subcaption}
\usepackage[title]{appendix}
\usepackage{makecell}
\usepackage{authblk}
\usepackage[sorting=none, style=numeric-comp, backend=bibtex]{biblatex}
\usepackage[left=1.00 in, right=1.00 in, top=1.00 in, bottom=1.00 in]{geometry}

\title{\LARGE \textbf{Development of a Weakly Compressible Solver for Incompressible Two-Phase Flows}}

\author{Ashley Melvin, J. C. Mandal}

\affil{Department of Aerospace Engineering, Indian Institute of Technology Bombay, Mumbai, Maharashtra, 400076, India}

\date{}

\begin{document}

\maketitle

\begin{abstract}
	In this paper, a novel fully-explicit weakly compressible solver is developed for solving incompressible two-phase flows. The two-phase flow is modelled by coupling the general pressure equation, momentum conservation equations and the conservative level set advection equation. A HLLC-type Riemann solver is proposed to evaluate the convective fluxes along with a simple, consistent and oscillation-free discretization for the non-conservative terms. The solver is tested against several two-phase flow problems for its robustness and adaptability on structured as well as unstructured meshes. 
\end{abstract}


\section{Introduction}
\label{sec:into}

Numerical methods for solving the incompressible Navier-Stokes equations are challenged by the absence of a pressure evolution equation. Traditional computational fluid dynamics (CFD) solvers, utilizing pressure-based methods \cite{patankar1983calculation, van1984enhancements}, address this by solving the pressure Poisson equation; a computationally intensive process with convergence issues, particularly in two-phase flow problems \cite{cerne2001coupling,sussman2007sharp}. An alternative is the density-based approach, typically used in compressible flow solvers. A widely adopted method in this framework is Chorin's artificial compressibility (AC) method \cite{chorin1967numerical}, initially developed for steady-state problems. Although it has been extended to handle unsteady flows \cite{gaitonde1998dual, malan2002improveda, malan2002improvedb, nithiarasu2003efficient}, application of AC method in such cases requires dual time-stepping. This procedure involves excessive iteration in pseudo-time, to ensure a divergence-free velocity field, before marching in real time, rendering the method computationally expensive.

Recently, weakly compressible (WC) models have emerged as a more efficient alternative for incompressible flow simulations. These models derive a pressure evolution equation by taking the incompressible limit of compressible flow equations \cite{ansumali2005thermodynamic}, allowing fully-explicit algorithms that eliminate the need for sub-iterations. This leads to the development of scalable algorithms, particularly desirable in large-scale incompressible flow problems \cite{mirin1999very}. Furthermore, WC models, like the kinetically reduced local Navier-Stokes (KRLNS) equation \cite{borok2007kinetically}, entropically damped artificial compressibility (EDAC) method \cite{clausen2013entropically}, and the general pressure equation (GPE) \cite{toutant2017general}, can leverage advanced compressible flow algorithms \cite{pan2022high}. Among the WC models, GPE-based algorithms have shown superior computational efficiency \cite{toutant2018numerical, shi2020simulations, pan2022high}, as it lacks the superficial terms present in other WC models, without compromising the accuracy of the solver \cite{toutant2018numerical}.

The WC models have been tested on single phase flows extensively \cite{borok2007kinetically,clausen2013edac, delorme2017simple,toutant2018numerical,kajzer2018application,shi2020simulations,pan2022high}. However, the studies on two-phase flows using WC models are relatively scarce. One of the earliest studies \cite{matsushita2019weakly} considers a low Mach limit of the Euler equation along with the conservative Allen-Cahn equation \cite{chiu2011conservative} to develop a WC two-phase model. Another WC model was studied by Kajzer and Pozorski \cite{kajzer2020weakly} in which the EDAC flow model was coupled with the Cahn-Hillard phase-field model \cite{fakhari2017improved}. The model, however, required special treatment at fluid interfaces where the pressure evolution equation was modified. Several limitations of this WC model were resolved in a later work by the same authors \cite{kajzer2022weakly}. Yang and Aoki developed a WC solver using evolving pressure projection method on staggered \cite{yang2021weakly} as well as collocated \cite{yang2022momentum} grids with phase-field and volume of fluid \cite{hirt1981volume} methods respectively. The pressure evolution equation used in these studies is the GPE without the pressure diffusion term which aids in damping the acoustic waves. The authors argue that the evolving pressure projection damps the acoustic waves rendering the pressure diffusion term superficial. A similar evolving pressure projection method was also used in \cite{matsushita2021gas}.

Almost all of the existing WC two-phase flow algorithms solve the pressure, momentum and interface advection equations in a decoupled manner. As demonstrated in \cite{yang2014upwind}, solving the equations simultaneously removes any lag between the solution variables, which may be more conspicuous in fully-explicit algorithms, when marched in time. Furthermore, despite the possibility of adopting Riemann solvers, only a handful of Riemann solvers for WC two-phase models are reported in literature \cite{li2022simplified}. The ability of Riemann solvers to model the fluid interface as a contact wave along with the accurate jump conditions in pressure, velocity and phase function across the waves make them an exemplary choice to simulate two-phase flows \cite{toro2013riemann}. The robustness of Riemann solvers to accurately model incompressible two-phase flows has already been demonstrated using AC methods \cite{yang2014upwind,parameswaran2019novel,bhat2019contact,bhat2022improved}, opening avenues for extending these techniques to WC methods.

In the present work, a WC model using GPE coupled with the conservative level set (CLS) method \cite{olsson2005conservative, olsson2007conservative} is used to simulate incompressible two-phase flows. The CLS method has been adopted to model interface evolution owing to its superior mass conservation capability \cite{olsson2005conservative} when compared to the more popular signed-distance based level set method \cite{osher1988fronts}. The volume of fluid (VOF) method \cite{hirt1981volume}, another common interface capturing technique, while possessing good mass conservation property has a computationally intensive interface reconstruction procedure. Additionally, computation of interface normals and gradients can be erroneous due to the discontinuous nature of the volume fraction, across the fluid interface, which can lead to numerical challenges in the VOF method. The CLS method models the phase fraction using a smooth hyperbolic tangent function which in turn smoothens the jump in material properties (density, viscosity etc.) across the interface. Furthermore, in the CLS framework, the material properties are related to the CLS function through a direct linear expression, which simplifies the formulation and analysis of the HLLC (Harten-Lax-van Leer with Contact) \cite{toro1994restoration} Riemann solver, developed in the present work. The WC model used in this study leads to a non-conservative system. Path-conservative schemes \cite{pares2006numerical,castro2008many} based on the theoretical work in \cite{dal1995definition} are generally used to compute the solution of non-conservative hyperbolic problems. However, in the present work, the non-conservative products are discretized using a non-oscillatory scheme \cite{abgrall1996prevent, saurel1999multiphase, saurel2018diffuse} for its simplicity. The solver is validated against benchmark results from several incompressible two-phase flow problems, available in literature, to demonstrate its robustness. Simulations are performed on structured as well as unstructured grids to attest the adaptability of the proposed algorithm.

The mathematical formulation of the WC two-phase model used in the study is presented in section \ref{sec:math}, followed by the numerical methods for discretizing the model in section \ref{sec:num}. Section \ref{sec:res} discusses the results and inferences from the numerical experiments. Finally, the conclusions drawn from the present work are outlined in section \ref{sec:conc}.

\section{Mathematical model}\label{sec:math}

The present work aims to simulate unsteady incompressible viscous flows of two immiscible, Newtonian fluids using the weakly compressible (WC) general pressure equation (GPE). For simplicity, an isothermal, laminar two-dimensional flow is assumed.

\subsection{Governing equations}

In the WC framework, the continuity equation is replaced with a pressure evolution equation. The pressure evolution equation used in the present work is the GPE \cite{toutant2018numerical}
\begin{equation}\label{eq:math_gpe_nonC}
	\dfrac{\partial p}{ \partial t} + \beta \rho \left( \dfrac{\partial u}{\partial x} + \dfrac{\partial v}{\partial y} \right) = \dfrac{\partial}{\partial x}\left( \dfrac{\mu}{\rho} \dfrac{\partial p}{\partial x}\right) +  \dfrac{\partial}{\partial y}\left( \dfrac{\mu}{\rho} \dfrac{\partial p}{\partial y}\right)
\end{equation}
where $p$ is the pressure and ${\mathbf{v}} \equiv (u,v)$ is the velocity vector. The parameter $\beta$ represents the artificial compressibility (AC) parameter, while the material properties $\rho$ and $\mu$ are the density and dynamic viscosity respectively. The derivation of GPE from the compressible energy equation is detailed in Toutant's work \cite{gpe1}. The form of GPE in \eqref{eq:math_gpe_nonC} consists of non-conservative terms. To ensure consistency with the momentum \eqref{eq:math_mom} and interface evolution \eqref{eq:math_cls} equations, which contain conservative terms, GPE \eqref{eq:math_gpe_nonC} is rewritten as
\begin{equation}\label{eq:math_gpe}
	\dfrac{\partial (p/\beta)}{ \partial t} +  \dfrac{\partial (\rho u)}{\partial x} + \dfrac{\partial (\rho v)}{\partial y} - \left( u\dfrac{\partial \rho}{\partial x} + v\dfrac{\partial \rho}{\partial y} \right)  = \dfrac{\partial}{\partial x}\left( \dfrac{\mu}{\beta\rho} \dfrac{\partial p}{\partial x}\right) +  \dfrac{\partial}{\partial y}\left( \dfrac{\mu}{\beta\rho} \dfrac{\partial p}{\partial y}\right)
\end{equation}
Writing the pressure evolution equation in the above non-conservative form ensures consistent coupling with the other governing equations. A similar transformation is used while writing the interface advection equation \eqref{eq:math_ls} as well.

The pressure evolution equation \eqref{eq:math_gpe} is combined with the momentum conservation equations
\begin{subequations}\label{eq:math_mom}
	\begin{align}
		\dfrac{\partial (\rho u)}{\partial t} + \dfrac{\partial (\rho u^2 + p)}{\partial x} + \dfrac{\partial (\rho uv)}{\partial y} &= \dfrac{\partial}{\partial x}\left( 2 \mu \dfrac{\partial u}{\partial x}\right) +  \dfrac{\partial}{\partial y}\left\{ \mu \left(\dfrac{\partial u}{\partial y} + \dfrac{\partial v}{\partial x} \right) \right\} \\
		\dfrac{\partial (\rho v)}{\partial t} + \dfrac{\partial (\rho uv)}{\partial x} + \dfrac{\partial (\rho v^2 + p)}{\partial y} &= \dfrac{\partial}{\partial x}\left\{ \mu \left(\dfrac{\partial u}{\partial y} + \dfrac{\partial v}{\partial x} \right) \right\} + \dfrac{\partial}{\partial y}\left( 2 \mu \dfrac{\partial v}{\partial y}\right)
	\end{align}
\end{subequations}
and the interface advection equation
\begin{equation}\label{eq:math_ls}
	\dfrac{\partial \psi}{\partial t} + \dfrac{\partial (u\psi)}{\partial x} + \dfrac{\partial (v\psi)}{\partial y} = \psi \left( \dfrac{\partial u}{\partial x} + \dfrac{\partial v}{\partial y} \right)
\end{equation}
where $\psi$ is the level set function. In truly incompressible models, the term on the right-hand side of the level set advection equation \eqref{eq:math_ls} drops to zero. Despite the relaxed divergence-free velocity constraint in WC models, which can lead to non-trivial divergence of velocity field, the term on the right-hand side is neglected in the present study for the reasons reported in \cite{yang2022momentum}. Firstly, the compressibility can be considered negligible at sufficiently low Mach numbers, dictated by the choice of the AC parameter $\beta$. Secondly, in truly incompressible models, the continuity equation serves as the mass conservation equation. However, the continuity equation is replaced with GPE in the present model. Therefore, the conservative level set equation serves as the equation of mass conservation in this model \cite{yang2022momentum}.
\begin{equation}\label{eq:math_cls}
	\dfrac{\partial \psi}{\partial t} + \dfrac{\partial (u\psi)}{\partial x} + \dfrac{\partial (v\psi)}{\partial y} = 0
\end{equation}
A hyperbolic tangent function \cite{olsson2005conservative} is considered as the conservative level set function $\psi$
\begin{equation}\label{eq:math_cls_func}
	\psi(x,y,t) = \dfrac{1}{2}\left\{ \tanh\left(\frac{\phi(x,y,t)}{2\varepsilon}\right) + 1\right\} = \dfrac{1}{1 + e^{-\frac{\phi(x,y,t)}{\varepsilon}}}
\end{equation}
where $\epsilon$ is a mesh dependent parameter that dictates the width of the smooth transition region between the two fluids and $\phi$ is the signed distance function
\begin{equation}
	\phi(x,y,t) =
	\begin{cases}
		+d, & \text{inside fluid 1} \\
		-d, & \text{inside fluid 2}
	\end{cases}
\end{equation}
with $d$ being the shortest distance from the fluid interface. For the level set field given by \eqref{eq:math_cls_func}, the fluid interface is represented by the $\psi = 0.5$ contour. The material properties such as density and viscosity are related to the level set field \eqref{eq:math_cls_func} through the following expression
\begin{equation}\label{eq:math_material}
	(\cdot) = (\cdot)_1 \psi + (\cdot)_2 (1 - \psi)
\end{equation}
Here $(\cdot)$ can be any material property (density $\rho$, dynamic viscosity $\mu$ etc.), and $(\cdot)_1$ and $(\cdot)_2$ are the corresponding properties of fluid 1 and 2 respectively.

The governing equations \eqref{eq:math_gpe}, \eqref{eq:math_mom} and \eqref{eq:math_cls} along with the gravitational and surface tension forces can be written in a compact form as
\begin{equation}\label{eq:math_compact}
	\dfrac{\partial \mathbf{U}}{\partial t} + \dfrac{\partial \mathbf{F}_c}{\partial x} + \dfrac{\partial \mathbf{G}_c}{\partial y} + \mathbf{B}_x \dfrac{\partial \mathbf{U}}{\partial x} + \mathbf{B}_y \dfrac{\partial \mathbf{U}}{\partial y} = \dfrac{\partial \mathbf{F}_d}{\partial x} + \dfrac{\partial \mathbf{G}_d}{\partial y} + \mathbf{F}_g + \nabla \cdot \mathbf{F}_s
\end{equation}
where
\begin{equation*}
	\begin{split}
		\mathbf{U} = 
		\begin{bmatrix}
			p/\beta \\
			\rho u \\
			\rho v \\
			\psi
		\end{bmatrix},\quad
		\mathbf{F}_c = 
		\begin{bmatrix}
			\rho u \\
			\rho u^2 + p \\
			\rho uv \\
			u \psi
		\end{bmatrix},\quad
		\mathbf{G}_c &=
		\begin{bmatrix}
			\rho v \\
			\rho uv \\
			\rho v^2 + p \\
			v \psi
		\end{bmatrix}, \quad
		\mathbf{B}_x =
		\begin{bmatrix}
			0 & 0 & 0 & -(\rho_1 - \rho_2)u \\
			0 & 0 & 0 & 0 \\
			0 & 0 & 0 & 0 \\
			0 & 0 & 0 & 0 
		\end{bmatrix}, \\
		\mathbf{B}_y =
		\begin{bmatrix}
			0 & 0 & 0 & -(\rho_1 - \rho_2)v \\
			0 & 0 & 0 & 0 \\
			0 & 0 & 0 & 0 \\
			0 & 0 & 0 & 0 
		\end{bmatrix},\quad
		\mathbf{F}_d &= 
		\begin{bmatrix}
			\frac{\mu}{\beta\rho} \frac{\partial p}{\partial x} \\
			2\mu \frac{\partial u}{\partial x} \\
			\mu \left( \frac{\partial u}{\partial y} + \frac{\partial v}{\partial x}\right)\\
			0
		\end{bmatrix},\quad 
		\mathbf{G}_d = 
		\begin{bmatrix}
			\frac{\mu}{\beta\rho} \frac{\partial p}{\partial y} \\
			\mu \left( \frac{\partial u}{\partial y} + \frac{\partial v}{\partial x}\right)\\
			2\mu \frac{\partial v}{\partial y}\\
			0
		\end{bmatrix}, \quad \\
		\mathbf{F}_g = 
		\begin{bmatrix}
			0\\
			\rho g_x \\
			\rho g_y \\
			0
		\end{bmatrix},\quad & \text{and}\quad
		\nabla \cdot \mathbf{F}_s = 
		\begin{bmatrix}
			0\\
			\nabla \cdot \mathbf{T}_s \\
			0
		\end{bmatrix}
	\end{split}
\end{equation*}
Here $\mathbf{U}$ is the vector of conserved variables, $\mathbf{F}_c$ and $\mathbf{G}_c$ are the convective fluxes in the $x$ and $y$ directions respectively. $\mathbf{B}_x$ and $\mathbf{B}_y$ are the coefficient matrices of the non-conservative products. The pressure diffusion term in GPE and viscous terms in momentum equations constitute the diffusive flux terms $\mathbf{F}_d$ and $\mathbf{G}_d$. The gravitational force $\mathbf{F}_g$ is considered as source term, where ${\mathbf{g}} \equiv (g_x,g_y)$ is the acceleration due to gravity. The tensors $\mathbf{F}_s$ and $\mathbf{T}_s$, associated with surface tension force, are detailed in subsequent section.

\subsubsection{Note on artificial compressibility parameter}

Replacing the continuity equation with a pressure evolution equation (GPE) \eqref{eq:math_gpe} introduces acoustic waves of finite speed into the \emph{truly} incompressible system with infinite speed of sound. In the present framework, the speed of sound of the system is dictated by the artificial compressibility (AC) parameter $\beta$ which is equal to the square of the speed of sound \cite{toutant2018numerical}. In GPE \eqref{eq:math_gpe}, as the AC parameter approaches infinity, the continuity equation with the divergence-free velocity constraint can be recovered. Therefore, to mimic an incompressible system, WC solvers would need an appropriately large AC parameter. However, a high AC parameter would impose severe time-step restrictions to the numerical simulations. The AC parameter, defined as
\begin{equation}
	\beta = \left( \dfrac{U_\text{max}}{M\!a}\right)^2
\end{equation}
where $U_\text{max}$ is the maximum expected magnitude of velocity in the flow, is dictated by the choice of Mach number $M\!a$. In the literature \cite{yang2022momentum, matsushita2019weakly, kajzer2020weakly, yang2021weakly, matsushita2021gas, kajzer2022weakly}, it is generally agreed that a Mach number less than 0.1 is adequate for WC models to simulate incompressible flows,a standard that the present work also adopts.

\subsection{Reinitialization of the level set field}

Due to the numerical errors in the algorithms used to advect the interface \eqref{eq:math_compact}, the level set function loses its property defined in \eqref{eq:math_cls_func} over time. The dissipative nature of the solver leads to the smearing of the fluid interface over time, necessitating a reinitialization procedure to restore this property. In the pioneering work on the conservative level set method \cite{olsson2005conservative}, an artificial compression-based reinitialization procedure was introduced \cite{harten1977artificial}, which was later improved in \cite{olsson2007conservative}. The reinitialization procedure has two main drawbacks \cite{parameswaran2023stable}: i) undesired interface movement, and ii) unphysical patch formation away from the interface. In the literature, several remedies have been proposed \cite{desjardins2008accurate,shukla2010interface,mccaslin2014localized,waclawczyk2015consistent,chiodi2017reformulation,shervani2018stabilized,parameswaran2023stable}, albeit at the expense of computational efficiency and/or conservation. In the present work, the reinitialization technique from the stabilized conservative level set (SCLS) method \cite{shervani2018stabilized} is adopted. In the SCLS method, the interface normal is estimated as
\begin{equation}\label{eq:math_interNormal_scls}
	\mathbf{n}_\psi = \dfrac{\nabla \psi}{\sqrt{|\nabla \psi|^2 + \varepsilon \exp\left(-\delta \varepsilon^2|\nabla \psi|^2 \right)}}
\end{equation}
where the tunable parameter $\delta = 10$, as recommended in \cite{shervani2018stabilized}. The magnitude of the interface normal estimated by \eqref{eq:math_interNormal_scls}, diminishes away from the fluid interface and to ensure the correct asymptotic behaviour throughout the domain, the reinitialization equation has an additional diffusion term (when compared to the reinitialization equation in \cite{olsson2007conservative}). The SCLS reinitialization equation is written as
\begin{equation}\label{eq:math_rein_scls}
	\dfrac{\partial \psi}{\partial \tau} + \nabla \cdot \left\{ \psi (1-\psi) \mathbf{n}_\psi \right\} = \nabla \cdot \left\{\varepsilon (\nabla \psi \cdot \mathbf{n}_\psi) \mathbf{n}_\psi \right\} + \nabla \cdot \left\{ (1 - |\mathbf{n}_\psi|^2) \varepsilon \nabla \psi \right\}
\end{equation}
Being in conservative form, the reinitialization equation \eqref{eq:math_rein_scls} can be discretized using standard finite volume techniques \cite{toro2013riemann}.

\section{Numerical method}\label{sec:num}

The presence of non-conservative terms in the present model precludes its treatment using standard finite volume methods. However, studies on multi-fluid compressible models \cite{saurel1999multiphase,saurel2001multiphase,andrianov2003simple} have led to the development of a simple and efficient methodology when dealing with non-conservative multiphase systems. In this approach, the conservative fluxes are discretized using the standard finite volume method \cite{toro2013riemann} and the non-conservative terms are treated according to a steady state constraint following the principle proposed by Abgrall \cite{abgrall1996prevent}. Numerical experiments \cite{karni1996hybrid, saurel1999multiphase, andrianov2003simple} on compressible multiphase flows have also indicated that this treatment of non-conservative system yields results with acceptably low conservation errors in the presence of weak to moderate shocks. Since the present study aims to simulate isothermal incompressible flows, strong shock considerations are not relevant, and a simple treatment of the non-conservative products, as in \cite{saurel1999multiphase}, is justified.

\subsection{Finite volume discretization}

The integral form of the governing equation \eqref{eq:math_compact} can be written as
\begin{equation}\label{eq:num_integral_1}
	\begin{split}
		\dfrac{\partial}{\partial t} \iint\displaylimits_\Omega \mathbf{U} \,d\Omega + \oint\displaylimits_\Gamma (\mathbf{F}_c n_x + \mathbf{G}_c n_y) \,d\Gamma & + \iint\displaylimits_\Omega \left(\mathbf{B}_x \dfrac{\partial \mathbf{U}}{\partial x} + \mathbf{B}_y \dfrac{\partial \mathbf{U}}{\partial y}\right) d\Omega \\ & = \oint\displaylimits_\Gamma (\mathbf{F}_d n_x + \mathbf{G}_d n_y) \,d\Gamma + \iint\displaylimits_\Omega \mathbf{F}_g \,d\Omega + \oint\displaylimits_\Gamma \mathbf{F_s}\cdot\mathbf{n} \,d\Gamma
	\end{split}
\end{equation}
where $d\Gamma$ is the infinitesimal length over the boundary $\Gamma$ of the control volume $\Omega$ and $\mathbf{n} \equiv (n_x,n_y)$ is the outward facing unit normal vector of $d\Gamma$.  

The treatment of the non-conservative terms follows the theory in \cite{saurel1999multiphase}. Only the aspects relevant to the present study are highlighted here and for details the readers are referred to \cite{saurel1999multiphase, saurel2018diffuse}. The assumption, at the outset, is that the spatial variation of the coefficient matrix within a control volume is negligible \cite{murrone2005five}. Therefore, the area integral of non-conservative terms can be replaced with line integral as
\begin{equation}
	\iint\displaylimits_\Omega \left(\mathbf{B}_x \dfrac{\partial \mathbf{U}}{\partial x} + \mathbf{B}_y \dfrac{\partial \mathbf{U}}{\partial y}\right) d\Omega \approx \mathbf{B}_x \oint\displaylimits_\Gamma \mathbf{U} n_x \,d\Gamma + \mathbf{B}_y \oint\displaylimits_\Gamma \mathbf{U} n_y \,d\Gamma
\end{equation}
Using the above approximation, the integral form of the governing equation \eqref{eq:num_integral_1} can be written as 
\begin{equation}\label{eq:num_integral_2}
	\begin{split}
		\dfrac{\partial}{\partial t} \iint\displaylimits_\Omega \mathbf{U} \,d\Omega + \oint\displaylimits_\Gamma (\mathbf{F}_c n_x + \mathbf{G}_c n_y) \,d\Gamma & + \mathbf{B}_x \oint\displaylimits_\Gamma \mathbf{U} n_x \,d\Gamma + \mathbf{B}_y \oint\displaylimits_\Gamma \mathbf{U} n_y \,d\Gamma \\ & = \oint\displaylimits_\Gamma (\mathbf{F}_d n_x + \mathbf{G}_d n_y) \,d\Gamma + \iint\displaylimits_\Omega \mathbf{F}_g \,d\Omega + \oint\displaylimits_\Gamma \mathbf{F_s}\cdot\mathbf{n} \,d\Gamma
	\end{split}
\end{equation}

\begin{figure}[ht!]
	\centering
	\includegraphics[width=0.6\textwidth]{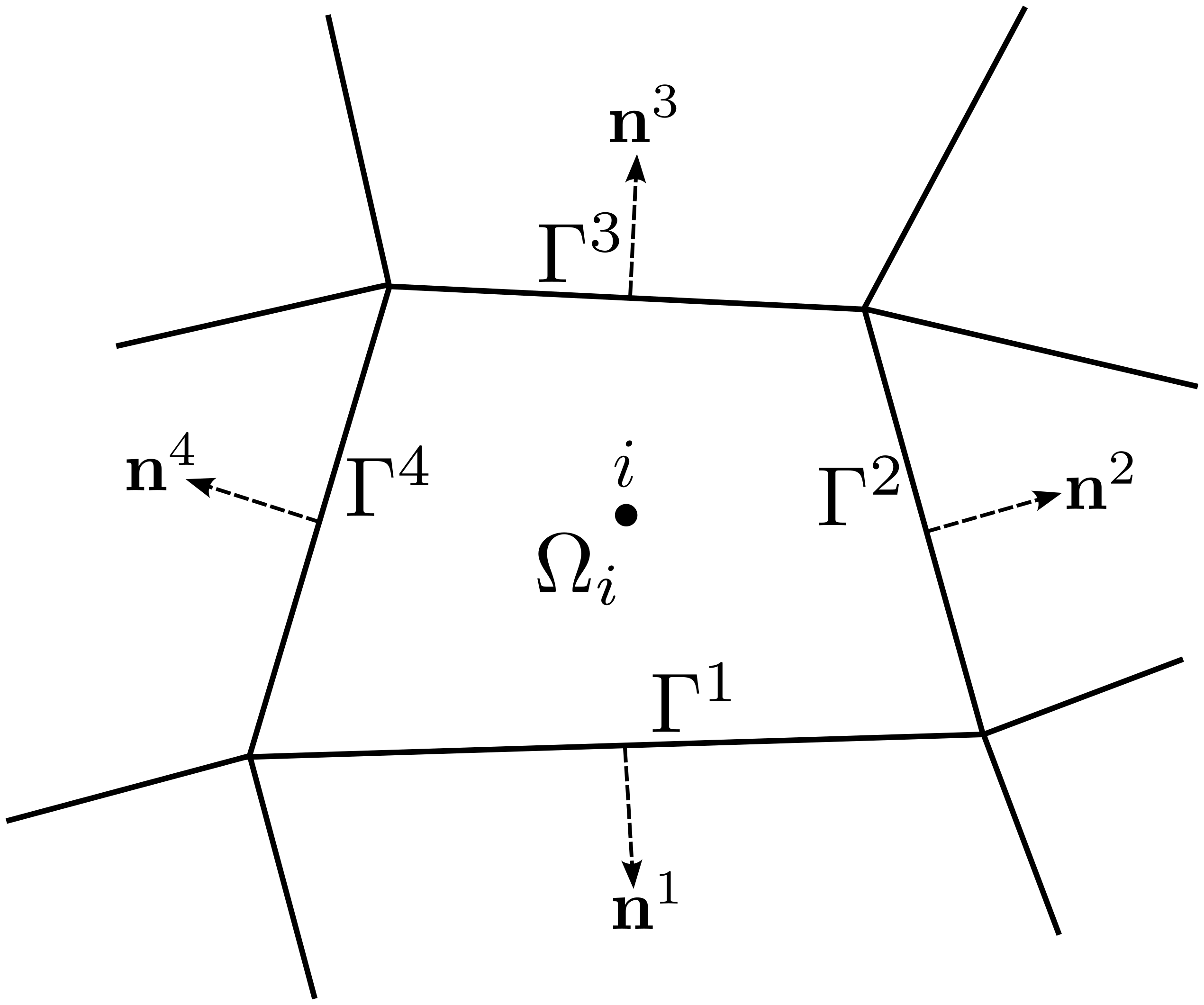}
	\caption{The $i^\text{th}$ finite volume cell with $M=4$ edges (cell interfaces). Here $\Omega_i$ denotes the area of the cell and $\Gamma^{(\cdot)}$ are the edge lengths. $\mathbf{n}^{(\cdot)}$ represent the $M$ outward facing normals of the $i^\text{th}$ cell.}
	\label{fig:num_fvmCell}
\end{figure}

The computational domain is discretized into a finite number of non-overlapping cells. A typical finite volume cell, used in this study, is shown in figure \ref{fig:num_fvmCell}. For the $i^\text{th}$ cell with $M$ boundary edges, spatially discretized form of \eqref{eq:num_integral_2} can be written as
\begin{equation}\label{eq:num_spaceDisc}
	\Omega_i \dfrac{\partial \overline{\mathbf{U}}_i}{\partial t} + \mathcal{R}(\overline{\mathbf{U}}_i) = 0
\end{equation}
where the residual $\mathcal{R}$ is defined as
\begin{equation*}\label{eq:num_residual}
	\begin{split}
		\mathcal{R}(\overline{\mathbf{U}}_i) &= \sum_{m=1}^M (\mathbf{F}^m_c n^m_x + \mathbf{G}^m_c n^m_y) \Gamma^m + \mathbf{B}_x(\overline{\mathbf{U}}_i)\sum_{m=1}^M \mathbf{U}^m n^m_x \Gamma^m + \mathbf{B}_y(\overline{\mathbf{U}}_i)\sum_{m=1}^M \mathbf{U}^m n^m_y \Gamma^m \\&- \sum_{m=1}^M (\mathbf{F}^m_d n^m_x + \mathbf{G}^m_d n^m_y) \Gamma^m - \Omega_i \mathbf{F}_g(\overline{\mathbf{U}}_i) - \sum_{m=1}^M \mathbf{F}_s^m \cdot \mathbf{n}^m \Gamma^m
	\end{split}
\end{equation*}
Here, $\overline{\mathbf{U}}_i$ is the cell averaged conserved variable vector for the $i^\text{th}$ cell with its area (volume in three-dimension) denoted as $\Omega_i$. The notation $(\cdot)^m$ refers to the vectors/tensors evaluated at the $m^\text{th}$ face of the finite volume cell, with its face length (area in three-dimension) represented by $\Gamma^m$. The gravitational source term is computed as the product of the cell area $\Omega_i$ and the cell averaged value of gravitational force $\mathbf{F}_g$. It is important to reiterate that the treatment of the non-conservative terms in \eqref{eq:num_integral_2} is an approximation and the solution variables $\mathbf{U}^m$ defined at the cell interfaces in \eqref{eq:num_spaceDisc} depend on the numerical flux formulation under a steady-state constraint \cite{abgrall1996prevent}.

\subsection{Computation of convective flux}

The convective fluxes in the spatially discretized governing equations \eqref{eq:num_spaceDisc} are defined at the cell interfaces. Finite volume discretization naturally introduces discontinuities at these interfaces leading to a Riemann problem at each interface. To resolve this, a Riemann solver is typically employed to compute the flux at the interface. In this work, a contact-preserving Riemann solver is developed to compute the convective fluxes. Since Riemann problems are generally associated with hyperbolic systems, it is essential to verify the hyperbolicity of the present model.

\subsubsection{Hyperbolicity in time}

For a control volume with its outward unit normal vector ${\mathbf{n}} \equiv (n_x,n_y)$, the inviscid terms in discretized equation \eqref{eq:num_spaceDisc} satisfy the rotational invariance property
\begin{equation}\label{eq:num_rotationalInv}
	\left\{ \mathbf{F}_c(\mathbf{U})n_x + \mathbf{G}_c(\mathbf{U})n_y \right\} + \left\{ \mathbf{U}\mathbf{B}_x(\mathbf{U})n_x + \mathbf{U}\mathbf{B}_y(\mathbf{U})n_y \right\} = \mathbf{T}^{-1} \left\{\mathbf{F}_c(\mathbf{TU}) + \mathbf{TU}\,\mathbf{B}_x(\mathbf{TU})\right\}
\end{equation}
Here rotation matrix $\mathbf{T}$ and its inverse $\mathbf{T}^{-1}$ are defined as
\begin{equation*}
	\mathbf{T} = 
	\begin{bmatrix}
		1 & 0 & 0 & 0 \\
		0 & n_x & n_y & 0 \\
		0 & -n_y & n_x & 0 \\
		0 & 0 & 0 & 1
	\end{bmatrix}\quad\text{and}\quad
	\mathbf{T}^{-1} = 
	\begin{bmatrix}
		1 & 0 & 0 & 0 \\
		0 & n_x & -n_y & 0 \\
		0 & n_y & n_x & 0 \\
		0 & 0 & 0 & 1
	\end{bmatrix}
\end{equation*}

\begin{figure}[ht!]
	\centering
	\includegraphics[width=0.6\textwidth]{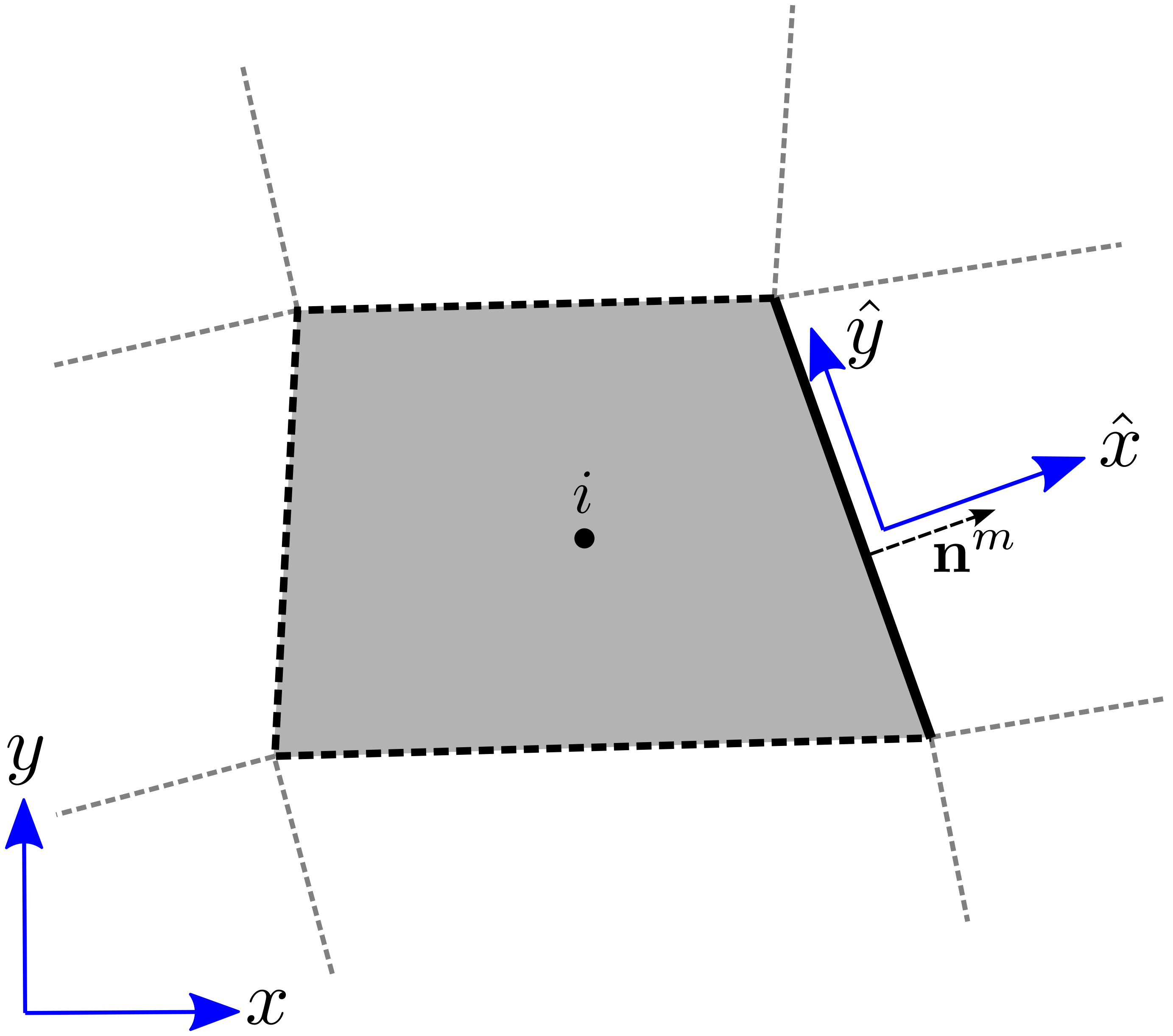}
	\caption{Locally rotated coordinate system $(\hat{x}, \hat{y})$ for the $m^\text{th}$ cell interface of the $i^\text{th}$ cell where $\mathbf{n}^m$ is its outward facing normal. Here $(x,y)$ represents the global Cartesian coordinate system.}
	\label{fig:num_rotationalInvariance}
\end{figure}

Consider a locally rotated coordinate system $(\hat{x}, \hat{y})$, schematically shown in figure \ref{fig:num_rotationalInvariance}, where the $\hat{x}$ is along the normal to the cell interface, and $\hat{y}$ tangential to it. In this coordinate system, the $\hat{x}$-split two-dimensional governing equations are sufficient for computing the convective flux as the other flux components do not contribute to the finite volume flux computation. Therefore, to compute the convective flux using a Riemann solver, the following system of governing equations is sufficient
\begin{equation}\label{eq:num_xSplit}
	\dfrac{\partial \hat{\mathbf{U}}}{\partial t} + \dfrac{\partial \mathbf{F}_c(\hat{\mathbf{U}})}{\partial \hat{x}} + \mathbf{B}_{\hat{x}}(\hat{\mathbf{U}}) \dfrac{\partial \hat{\mathbf{U}}}{\partial \hat{x}} = 0
\end{equation}
Here $\hat{\mathbf{U}} = \mathbf{TU} = \left[ p/\beta, \, \rho\hat{u}, \, \rho\hat{v}, \, \psi \right]^T$ with $\hat{u} = u n_x + v n_y$ and  $\hat{v} = -u n_y + v n_x$. The non-conservative equation \eqref{eq:num_xSplit} can be written in quasi-linear form as
\begin{equation}\label{eq:num_quasiLinear}
	\dfrac{\partial \hat{\mathbf{U}}}{\partial t} + \mathbf{A}(\hat{\mathbf{U}}) \dfrac{\partial \hat{\mathbf{U}}}{\partial \hat{x}} = 0
\end{equation}
where $\mathbf{A}(\hat{\mathbf{U}})$ is defined as
\begin{equation*}
	\mathbf{A}(\hat{\mathbf{U}}) = \dfrac{\partial \mathbf{F}_c(\hat{\mathbf{U}})}{\partial \hat{\mathbf{U}}} + \mathbf{B}_{\hat{x}}(\hat{\mathbf{U}})
\end{equation*}
Considering the relation \eqref{eq:math_material} between the level set function $\psi$ and material property namely, density $\rho$, the Jacobian of the convective flux vector $\mathbf{F}_c$ is given by
\begin{equation}\label{eq:num_jacobian}
	\dfrac{\partial \mathbf{F}_c(\hat{\mathbf{U}})}{\partial \hat{\mathbf{U}}} = 
	\begin{bmatrix}
		0 & 1 & 0 & 0 \\
		\beta & 2\hat{u} & 0 & -(\rho_1 - \rho_2)\hat{u}^2 \\
		0 & \hat{v} & \hat{u} & -(\rho_1 - \rho_2)\hat{u}\hat{v} \\
		0 & \dfrac{\psi}{\rho} & 0 & \dfrac{\rho_2 \hat{u}}{\rho}
	\end{bmatrix}
\end{equation}
The matrix
\begin{equation}\label{eq:num_nonConsMatrix}
	\mathbf{A}(\hat{\mathbf{U}}) = 
	\begin{bmatrix}
		0 & 1 & 0 & -(\rho_1 - \rho_2)\hat{u} \\
		\beta & 2\hat{u} & 0 & -(\rho_1 - \rho_2)\hat{u}^2 \\
		0 & \hat{v} & \hat{u} & -(\rho_1 - \rho_2)\hat{u}\hat{v} \\
		0 & \dfrac{\psi}{\rho} & 0 & \dfrac{\rho_2 \hat{u}}{\rho}
	\end{bmatrix}
\end{equation}
has four real eigenvalues
\begin{equation}\label{eq:num_eigenvalues}
	\lambda_1 = \hat{u}_\rho - \sqrt{\hat{u}^2_\rho + \beta},\quad \lambda_2 = \lambda_3 = \hat{u}\text{ and }\lambda_4 = \hat{u}_\rho + \sqrt{\hat{u}^2_\rho + \beta}
\end{equation}
where
\[ \hat{u}_\rho = \left(1 + \dfrac{\rho_2}{\rho}\right)\dfrac{\hat{u}}{2} \]
The corresponding right eigenvector matrix is given by
\begin{equation}\label{eq:num_eigenvectors}
	\mathbf{R} = 
	\begin{bmatrix}
		\mathbf{R}_1 & \mathbf{R}_2 & \mathbf{R}_3 & \mathbf{R}_4
	\end{bmatrix} = 
	\begin{bmatrix}
		1 & 0 & 0 & 1 \\
		\dfrac{\lambda_1}{\rho}\left(\dfrac{\rho\lambda_1 - \rho_2 \hat{u}}{\lambda_1 - \hat{u}}\right) & (\rho_1 - \rho_2)\hat{u} & 0 & \dfrac{\lambda_4}{\rho}\left(\dfrac{\rho\lambda_4 - \rho_2 \hat{u}}{\lambda_4 - \hat{u}}\right) \\
		\dfrac{\lambda_1 \hat{v}}{\lambda_1 - \hat{u}} & 0 & 1 & \dfrac{\lambda_4 \hat{v}}{\lambda_4 - \hat{u}} \\
		\dfrac{\psi}{\rho}\left(\dfrac{\lambda_1}{\lambda_1 - \hat{u}}\right) & 1 & 0 & \dfrac{\psi}{\rho}\left(\dfrac{\lambda_4}{\lambda_4 - \hat{u}}\right)
	\end{bmatrix}
\end{equation}
with four linearly independent columns, implying that the inviscid subsystem of governing equations \eqref{eq:math_compact} is \emph{hyperbolic in time}.

\subsubsection{Generalized Riemann invariant analysis}

A generalized Riemann invariant analysis \cite{toro2013riemann} of the intermediate waves, specifically the contact and shear waves, provides the precise jump conditions necessary for deriving the closed-form expressions for the intermediate states in the Riemann solver formulation. The generalized Riemann invariants across the contact wave $\mathbf{R}_2$ can be written as
\begin{equation}\label{eq:num_gri_contact}
	\dfrac{d(p/\beta)}{0} = \dfrac{d(\rho \hat{u})}{(\rho_1 - \rho_2)\hat{u}} = \dfrac{d(\rho \hat{v})}{0} = \dfrac{d \psi}{1}
\end{equation}
and across the shear wave $\mathbf{R}_3$ can be written as
\begin{equation}\label{eq:num_gri_shear}
	\dfrac{d(p/\beta)}{0} = \dfrac{d(\rho \hat{u})}{0} = \dfrac{d(\rho \hat{v})}{1} = \dfrac{d \psi}{0}
\end{equation}
Across both types of waves, the pressure $p$ remains constant. Mathematical manipulation, reveals that the normal velocity $\hat{u}$ is also constant across the intermediate waves. In contrast, a jump in tangential velocity $\hat{v}$ occurs across both the contact and shear waves. Additionally, the level set function remains constant across the shear wave but a jump is observed across the contact discontinuity confirming that the contact wave effectively models the fluid interface.

\subsubsection{HLLC Riemann solver}

\begin{figure}[hb!]
	\centering
	\includegraphics[width=0.7\textwidth]{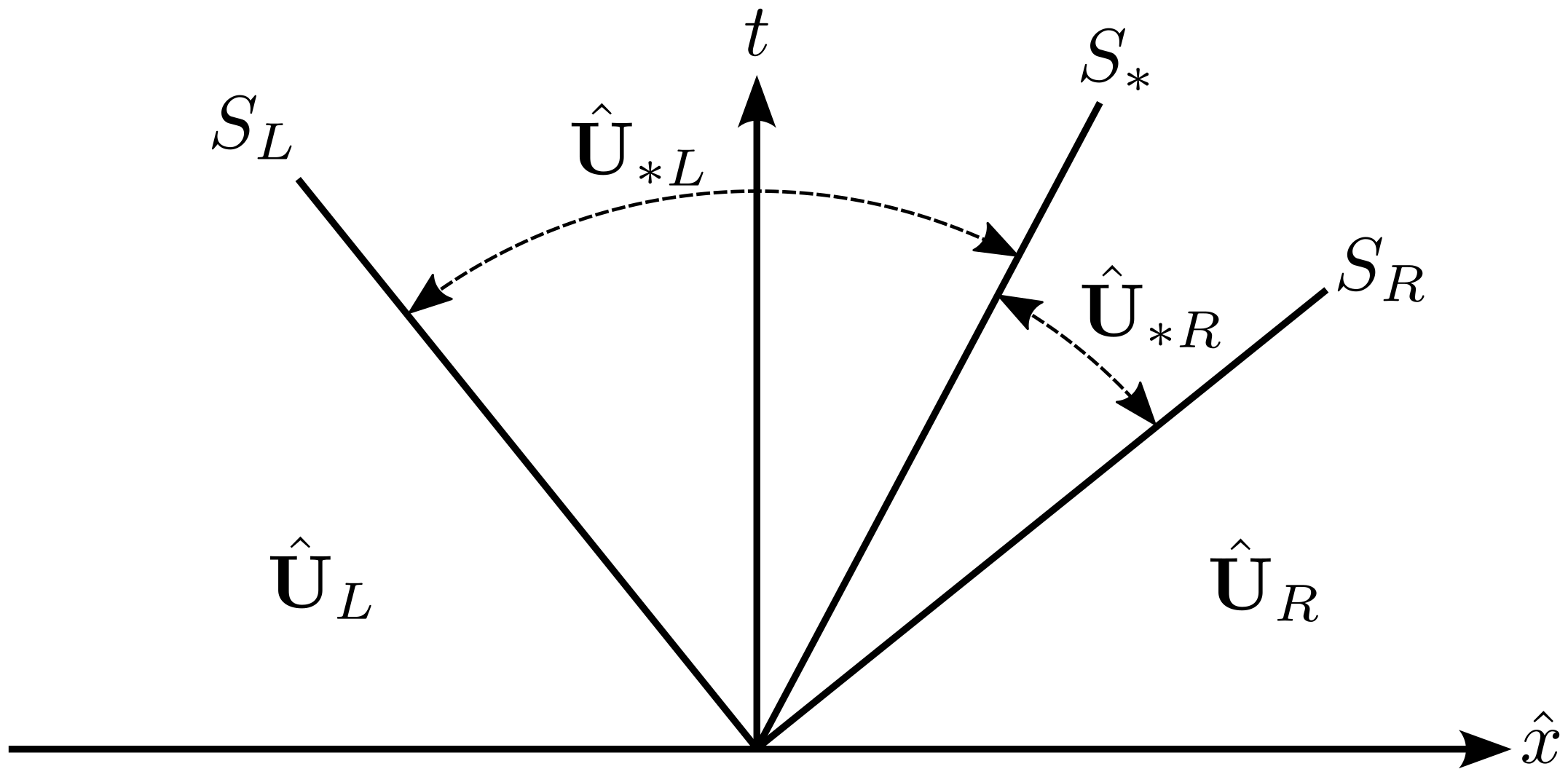}
	\caption{The wave structure of the HLLC Riemann solver.}
	\label{fig:num_hllc-waves}
\end{figure}

The present WC two-phase model, being hyperbolic in time, allows the developed of Riemann solvers. In this study, a contact-preserving HLLC Riemann solver \cite{toro1994restoration} is proposed to compute the convective flux. The wave structure, in a rotated co-ordinate system \eqref{eq:num_quasiLinear} (refer figure \ref{fig:num_rotationalInvariance}), in the HLLC formulation is shown in figure \ref{fig:num_hllc-waves}. As indicated in the figure, the HLLC solver assumes a three-wave model with a left and right running wave separated by an intermediate contact discontinuity. In the HLLC approximate Riemann solver, the numerical flux at the interface is given as
\begin{equation}\label{eq:num_hllcFlux}
	\mathbf{F}_c(\hat{\mathbf{U}}) =
	\begin{cases}
		\mathbf{F}(\hat{\mathbf{U}}_L), & S_L \geq 0 \\
		\mathbf{F}(\hat{\mathbf{U}}_L) + S_L(\hat{\mathbf{U}}_{*L} - \hat{\mathbf{U}}_L), & S_L < 0 \leq S_* \\
		\mathbf{F}(\hat{\mathbf{U}}_R) + S_R(\hat{\mathbf{U}}_{*R} - \hat{\mathbf{U}}_R), & S_* < 0 < S_R \\
		\mathbf{F}(\hat{\mathbf{U}}_R), & S_R \leq 0
	\end{cases}
\end{equation}
The intermediate states are obtained by solving the following equation
\begin{equation}\label{eq:num_hllc_totJump}
	\mathbf{F}_c(\hat{\mathbf{U}}_R) - \mathbf{F}_c(\hat{\mathbf{U}}_L) = S_L(\hat{\mathbf{U}}_{*L} - \hat{\mathbf{U}}_L) + S_*(\hat{\mathbf{U}}_{*R} - \hat{\mathbf{U}}_{*L}) + S_R(\hat{\mathbf{U}}_{R} - \hat{\mathbf{U}}_{*R})
\end{equation}
which represents the cumulative jump across the three waves.

\newtheorem*{remark}{Remark}
\begin{remark}
	The Rankine-Hugoniot jumps in \eqref{eq:num_hllc_totJump} consider only the conservative terms (convective fluxes). In path-conservative methods, a weak solution must satisfy the generalized Rankine-Hugoniot condition, which also includes the non-conservative products \cite{dal1995definition,gosse2001well}. In the current framework, the non-conservative terms are not explicitly present in the scheme for convective fluxes but they are implicitly included in the Riemann problem analyzed at each interface \cite{greenberg1996well}. The wave structure, considered for the formulation of the Riemann solver, is based on the quasi-linear form \eqref{eq:num_quasiLinear} that encompasses the non-conservative terms. However, the Riemann solver itself, as defined in \eqref{eq:num_hllcFlux}, is based solely on the conservative terms.
\end{remark}

Solving the equations \eqref{eq:num_hllc_totJump} component-wise, the intermediate variables $(\cdot)_{*L}$ and $(\cdot)_{*R}$ are obtained as
\begin{equation}\label{eq:num_hllcStar}
	\begin{split}
		(p/\beta)_{*L}  = (p/\beta)_{*R} = (p/\beta)_* &= \dfrac{(\rho \hat{u})_L - (\rho \hat{u})_R + S_R(p/\beta)_R - S_L(p/\beta)_L}{S_R - S_L} \\
		\hat{u}_{*L}  = \hat{u}_{*R} = \hat{u}_* = S_* &= \dfrac{S_L(\rho \hat{u})_L - S_R(\rho \hat{u})_R + (\rho \hat{u}^2+p)_R - (\rho \hat{u}^2+p)_L}{S_L\rho_L - S_R\rho_R + (\rho_1 - \rho_2)\left\{(\hat{u}\psi)_R - (\hat{u}\psi)_L\right\}} \\
		(\rho \hat{v})_{*L} &= \dfrac{S_L(\rho \hat{v})_L - (\rho \hat{u}\hat{v})_L}{S_L - S_*} \\
		(\rho \hat{v})_{*R} &= \dfrac{S_R(\rho \hat{v})_R - (\rho \hat{u}\hat{v})_R}{S_R - S_*} \\
		\psi_{*L} &= \dfrac{S_L \psi_L - (\hat{u}\psi)_L}{S_L - S_*} \\
		\psi_{*R} &= \dfrac{S_R \psi_R - (\hat{u}\psi)_R}{S_R - S_*} \\
	\end{split}
\end{equation}
The left and right wave speeds $S_L$ and $S_R$ are estimated as in \cite{davis1988simplified}
\begin{equation*}
	S_L = \min\{(\lambda_1)_L,(\lambda_1)_R\}\quad\text{and}\quad S_R = \max\{(\lambda_4)_L,(\lambda_4)_R\}
\end{equation*}
where $\lambda_1$ and $\lambda_4$ are given in \eqref{eq:num_eigenvalues}.

\subsubsection{Solution reconstruction}

\begin{figure}[ht!]
	\centering
	\includegraphics[width=0.8\textwidth]{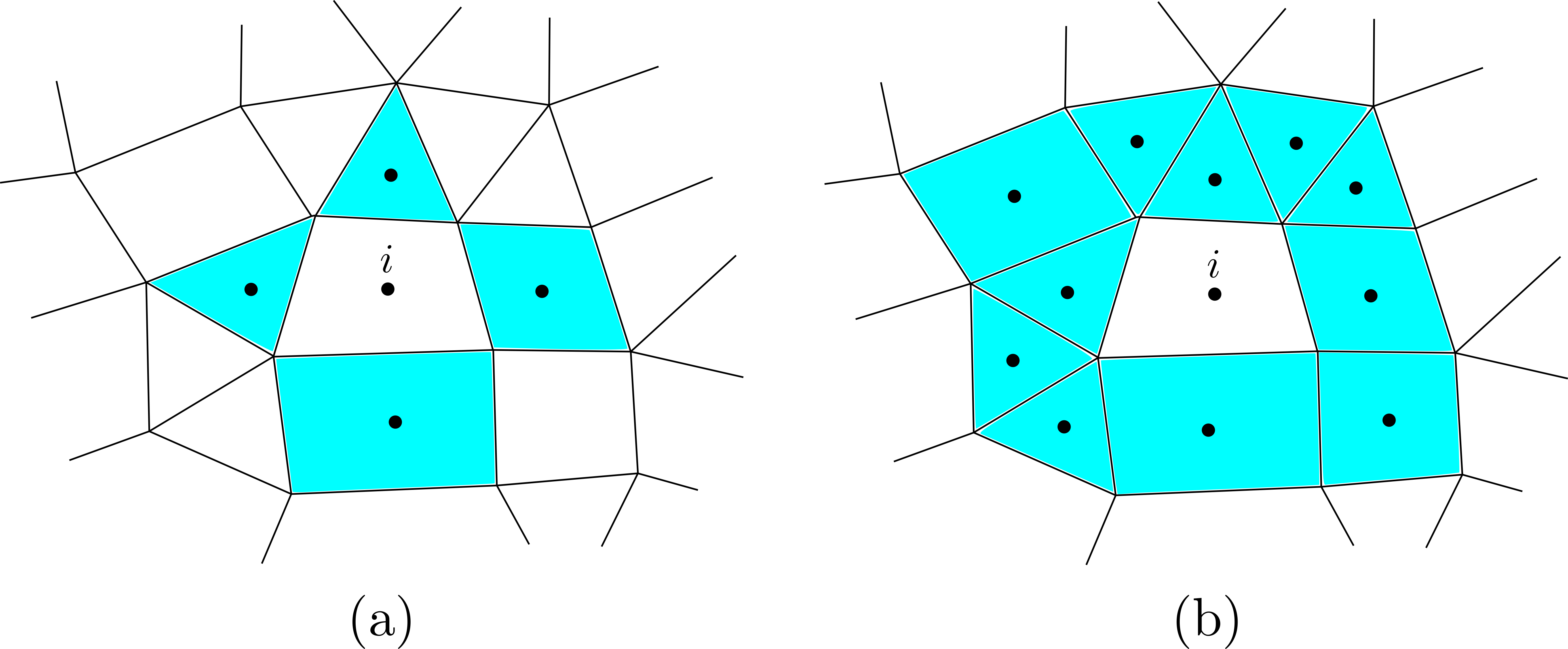}
	
	\caption{Stencils used for least square gradient computation at the $i^\text{th}$ cell centre: (a) cell edge based neighbors and (b) cell vertex based neighbors.}
	\label{fig:num_lsqNeighs}
\end{figure}

The convective fluxes at the cell interfaces, determined by the Riemann solver \eqref{eq:num_hllcFlux}, depend on the locally rotated values of the left and right states. Approximating these states by the corresponding cell centre values yields a first-order accurate scheme, which may not be sufficient in multiphase flow solvers where higher-order accuracy is crucial. In this study, the primitive variables namely, pressure, velocity and level set function are reconstructed at the cell interfaces using weighted least squares technique \cite{watkins2004fundamentals, barth1989design} with solution dependent weights \cite{mandal2008link, mandal2011high, mandal2015genuinely}. In the solution dependent weighted least squares (SDWLS) method, the gradient obtained at the cell centre is the \emph{limited} gradient, eliminating the need for any additional limiting procedures to obtain non-oscillatory solution. For structured Cartesian grids, a stencil involving cell interface neighbors is sufficient for gradient estimation. However, in \emph{truly} unstructured grids, face neighbor stencil may not provide sufficiently accurate results and may occasionally lead to instabilities \cite{haider2009stability, zangeneh2017reconstruction}. Therefore, in unstructured grids, a cell node/vertex based stencil is used for (limited) gradient estimation. The two types of stencils used for gradient estimation in the present work is shown in figure \ref{fig:num_lsqNeighs}. Readers are referred to \cite{parameswaran2023conservative} for detailed derivation of the over-determined system used for gradient estimation. In the present work, the over-determined system is solved using QR factorization via modified Gram-Schmidt (MGS) \cite{watkins2004fundamentals} process.

\subsection{Computation of non-conservative products}

The discretization of the non-conservative terms, as previously outlined, is based on the convective flux formulation, under a steady-state constraint. For multiphase flows, this constraint, derived from the principle by Abgrall \cite{abgrall1996prevent}, requires that \emph{a uniform constant pressure and velocity throughout the domain should remain unaltered over time}. By setting the spatial derivatives of pressure $p$ and velocity $(u,v)$, in the governing equations \eqref{eq:math_gpe} and \eqref{eq:math_mom} to zero, it can be shown that continuous pressure and velocity fields satisfy the Abgrall's principle, i.e., their temporal derivatives remain zero.

When this constraint, is applied to the space discretized equation \eqref{eq:num_spaceDisc}, it yields a discretization for the non-conservative terms that remains consistent with the numerical flux formulation. Since the non-conservative terms appear only in the pressure evolution equation, the first component of the discretized equations \eqref{eq:num_spaceDisc} is considered. For simplicity, the pressure diffusion terms are neglected in this analysis, which is also justified as the spatial derivatives of pressure are zero under the uniform pressure constraint. Invoking rotational invariance property \eqref{eq:num_rotationalInv} and the relation between level set function and density \eqref{eq:math_material}, the space discretized pressure evolution equation can be written as
\begin{equation}\label{eq:num_pressureSpaceDisc}
	\Omega_i \dfrac{\partial \overline{(p/\beta)}_i}{\partial t} + \sum_{m=1}^M \left\{ \left[\mathbf{F}_c(\hat{\mathbf{U}}^m) \right]_{11} - \hat{\overline{u}}^m_i\,\rho^m_\text{NC} \right\} \Gamma^m = 0
\end{equation}
where $\left[\mathbf{F}_c(\hat{\mathbf{U}}) \right]_{11}$ denotes the first component of the flux column vector and $\rho_\text{NC}$ denotes the non-conservative discretization that is to be determined. The steady-state constraint on pressure would result in the following equality at every cell interface
\begin{equation}
	\hat{\overline{u}}^m_i\,\rho^m_\text{NC} = \left[\mathbf{F}^m_c(\hat{\mathbf{U}}) \right]_{11}\quad\forall\,m
\end{equation}
Substituting the formulation of the HLLC flux \eqref{eq:num_hllcFlux} and using the uniform pressure and velocity constraint
\begin{equation*}
	\hat{\overline{u}}^m_i = \text{constant} \quad\text{and}\quad \overline{p}_i = \text{constant} \quad \forall\, i, m
\end{equation*}
the discretization for the non-conservative product is obtained as
\begin{equation}\label{eq:num_nonconv_rho}
	\rho^m_\text{NC} =
	\begin{cases}
		\rho_L, & S_L \geq 0 \\
		\dfrac{\rho_L S_R - \rho_R S_L}{S_R - S_L}, & S_L < 0 < S_R \\
		\rho_R, & S_R \leq 0
	\end{cases}
\end{equation}

\subsection{Computation of diffusive flux}

\begin{figure}[h]
	\centering
	\includegraphics[width=0.5\textwidth]{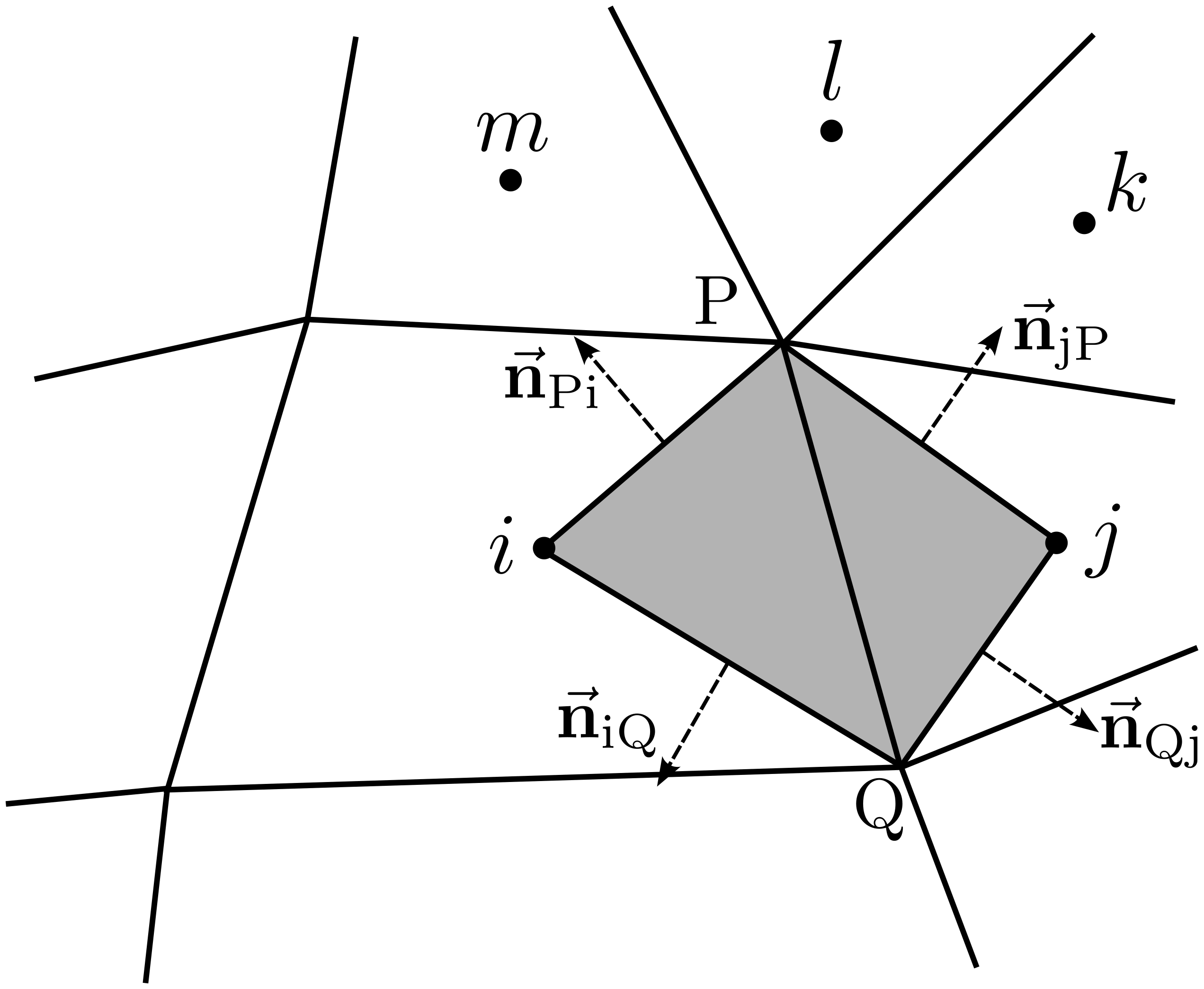}
	\caption{Estimation of diffusive fluxes. Green-Gauss approach along a Coirier diamond path around the edge PQ; counter-clockwise boundary of the shaded region.}
	\label{fig:num_coirierDiamond}
\end{figure}

Computation of diffusive fluxes $\mathbf{F}_d$ and $\mathbf{G}_d$ in \eqref{eq:math_compact} requires the estimation of gradients of pressure and velocity at the cell interfaces. The present work adopts the Green-Gauss approach to calculate the variable derivatives at the cell interfaces. In this approach, the Green-Gauss divergence theorem \cite{kreyszig2007advanced} is applied along a closed diamond (Coirier diamond \cite{coirier1994adaptively}) path connecting the adjacent cell centres and vertices of a particular edge as shown in figure \ref{fig:num_coirierDiamond}.

Consider a scalar field $f$ whose derivatives at the edge PQ needs to be computed. The gradient of the field $f$, computed using the Green-Gauss approach is
\begin{equation}\label{eq:num_greenGauss}
	\nabla f = \dfrac{1}{\Omega_\text{iQjP}}\left(f_\text{iQ} \vec{\mathbf{n}}_\text{iQ}\Gamma_\text{iQ} + f_\text{Qj} \vec{\mathbf{n}}_\text{Qj}\Gamma_\text{Qj} + f_\text{jP} \vec{\mathbf{n}}_\text{jP}\Gamma_\text{jP} + f_\text{Pi} \vec{\mathbf{n}}_\text{Pi}\Gamma_\text{Pi}\right)
\end{equation}
where $\Omega_\text{iQjP}$ is the area of the diamond (shaded region in figure \ref{fig:num_coirierDiamond}). $\vec{\mathbf{n}}_{(\cdot)}$ and $\Gamma_{(\cdot)}$ are the outward normal and length of a given edge $(\cdot)$ of the diamond respectively. The field values at the edges of the diamond is estimated by simply averaging the values at its end points, i.e.
\begin{equation*}
	f_\text{iQ} = \dfrac{f_\text{i} + f_\text{Q}}{2}
\end{equation*}
The cell centre field values $f_\text{i}$ and $f_\text{j}$ are known from the solution field. However, the field values at the cell vertices $f_\text{P}$ and $f_\text{Q}$ need to be computed. In this study, the cell vertex values are taken as the area weighted average of the neighbouring cell centre values. For a cell structure shown in figure \ref{fig:num_coirierDiamond}, the field value at vertex P is calculated as
\begin{equation*}
	f_\text{P} = \dfrac{\sum_{n \in N} \Omega_n f_n}{\sum_{n \in N} \Omega_n} \quad\text{where}\quad N = \{i,j,k,l,m\}
\end{equation*}
Finally, the field values at the cell interfaces, required for the computation of material properties in the diffusive flux terms, are calculated by averaging the values at the four vertices of the Coirier diamond around it.

\subsection{Computation of surface tension force}

In two-phase flows, the surface tension force, acting at the fluid interface, needs to be modelled accurately. The two prevalent methods used to model surface tension forces are the continuum surface force (CSF) \cite{brackbill1992continuum} and continuum surface stress (CSS) \cite{lafaurie1994modelling} models. The CSF model computes the volumetric surface tension force which can be taken as a source term in the governing equation. However, it requires the explicit computation of interface curvature which is prone to instabilities due to the ill-conditioned interface normals away from the interface \cite{shervani2018stabilized}. Therefore, in the present work the CSS model is employed to compute the surface tension tensor $\mathbf{T}_s$ in \eqref{eq:math_compact}, which can be described as
\begin{equation}\label{eq:num_css}
	\mathbf{T}_s = \sigma \left( |\nabla \psi| \mathbf{I} - \dfrac{\nabla \psi \otimes \nabla \psi}{|\nabla \psi|}  \right)
\end{equation}
Here $\sigma$ is the surface tension coefficient and $\mathbf{I}$ is the unit tensor. As indicated in the space discretized equation \eqref{eq:num_spaceDisc}, the surface tension forces $\mathbf{F}_s$ are computed at the cell interfaces. The gradient of the level set function $\nabla \psi$ in \eqref{eq:num_css}, at the cell interfaces, are computed along with the gradients required for the diffusive fluxes as detailed previously.

\subsection{Discretization of the reinitialization equation}

To maintain the sharpness of the fluid interface, the level set field is reinitialized after every few time steps. In the present work, the reinitialization method of stabilized conservative level set (SCLS) method \eqref{eq:math_rein_scls} as described in \cite{shervani2018stabilized} is implemented. The finite volume space discretized form of \eqref{eq:math_rein_scls} can be written as
\begin{equation}\label{eq:num_reinSpaceDisc}
	\Omega_i\dfrac{\partial \overline{\psi}_i}{\partial \tau} + \sum_{m=1}^{M} \mathbf{f}_c^m \cdot \vec{\mathbf{n}}^m \Gamma^m = \sum_{m=1}^{M} \mathbf{f}_d^m \cdot \vec{\mathbf{n}}^m \Gamma^m
\end{equation}
where the $\mathbf{f}_c^m$ and $\mathbf{f}_d^m$ compressive and diffusive fluxes respectively at the $m^\text{th}$ cell interface. The compressive flux is approximated as
\begin{equation*}
	\mathbf{f}_c^m = \dfrac{ \left\{ \psi (1-\psi) \mathbf{n}^0_\psi \right\}_L + \left\{ \psi (1-\psi) \mathbf{n}^0_\psi \right\}_R }{2}
\end{equation*}
where $(\cdot)_L$ and $(\cdot)_R$ are the cell-centred values at left and right cells of the $m^\text{th}$ cell interface. The cell-centered interface normal $\mathbf{n}^0_\psi$ is estimated using weighted least squares technique with distance based weights \cite{barth1989design}. The diffusive flux in the reinitialization equation is estimated as 
\begin{equation*}
	\mathbf{f}_d^m = \left\{ \varepsilon (\nabla \psi \cdot \mathbf{n}^0_\psi) \mathbf{n}^0_\psi \right\}_m + \left\{ (1 - |\mathbf{n}^0_\psi|^2) \varepsilon \nabla \psi \right\}_m
\end{equation*}
Here the gradients $\nabla \psi$ and interface normal $\mathbf{n}^0_\psi$ at the $m^\text{th}$ cell interface is estimated using a Green-Gauss technique similar to the one detailed for viscous flux estimation. Note that the interface normals $\mathbf{n}^0_\psi$ (at the cell centre and cell interface) \eqref{eq:math_interNormal_scls} are computed before the pseudo-time iterations and are not updated during the reinitialization step as recommended in \cite{olsson2005conservative}. A three stage strong stability preserving Runge-Kutta (SSP-RK) method \cite{gottlieb2009high} is used to iterate in pseudo-time until convergence. The mesh size dependent parameter $\varepsilon$ is taken as 
\begin{equation}\label{eq:num_epsilon}
	\varepsilon = \dfrac{h^{1-d}}{2}
\end{equation}
where $h$ is the average of the local cell sizes defined in \eqref{eq:num_cellSize}. The parameter $d \in [0,1)$ and as recommended in \cite{olsson2007conservative}, $d$ is taken as $0.1$ for good convergence. The reinitialization procedure \eqref{eq:num_reinSpaceDisc} reaches steady-state in only a few iterations \cite{olsson2007conservative} with the pseudo-time step $\Delta \tau$ given by
\begin{equation*}
	\Delta \tau = 2C_\tau \min_i\{h_i^{1+d}\}
\end{equation*}
For stability, the parameter $C_\tau$ is taken less than $0.25$ \cite{olsson2005conservative}.

\subsection{Temporal discretization}

As previously mentioned, the algorithm developed in the present work is fully-explicit. A three stage strong stability preserving Runge-Kutta (SSP-RK) method \cite{gottlieb2009high} is used to discretize the time derivatives. In the three stage SSP-RK method, the solution variable at $n^\text{th}$ time level $\overline{\mathbf{U}}^n$ is updated to $\overline{\mathbf{U}}^{n+1}$ for the $i^\text{th}$ cell as
\begin{equation}\label{eq:num_ssprk3}
	\begin{split}
		\overline{\mathbf{U}}^1_i &= \overline{\mathbf{U}}^n_i - \dfrac{\Delta t}{\Omega_i}\mathcal{R}\left( \overline{\mathbf{U}}^n_i \right) \\
		\overline{\mathbf{U}}^2_i &= \dfrac{3}{4}\overline{\mathbf{U}}^n_i + \dfrac{1}{4}\overline{\mathbf{U}}^1_i - \dfrac{1}{4}\dfrac{\Delta t}{\Omega_i}\mathcal{R}\left( \overline{\mathbf{U}}^1_i \right) \\
		\overline{\mathbf{U}}^{n+1}_i &= \dfrac{1}{3}\overline{\mathbf{U}}^n_i + \dfrac{2}{3}\overline{\mathbf{U}}^2_i - \dfrac{2}{3}\dfrac{\Delta t}{\Omega_i}\mathcal{R}\left( \overline{\mathbf{U}}^2_i \right)
	\end{split}
\end{equation}
where $\overline{\mathbf{U}}^1$ and $\overline{\mathbf{U}}^2$ are the solution variables at the intermediate stages. The global time-step $\Delta t$ is restricted by the stability requirements of the scheme. The time-step for the $i^\text{th}$ cell is given as
\begin{equation*}
	\Delta t_i = \min\{\Delta t^c_i,\Delta t^v_i,\Delta t^g_i,\Delta t^s_i\}
\end{equation*}
where the $\Delta t^c_i,\Delta t^v_i,\Delta t^g_i,\Delta t^s_i$ are the maximum allowable time-steps due to convective, viscous, gravitational and surface tension terms respectively.
These time-step restrictions are taken as \cite{blazek2015computational,mavriplis1990multigrid,swanson1992effective,brackbill1992continuum,parameswaran2019novel}
\begin{equation*}
	\begin{split}
		\Delta \tau^c_i \leq \dfrac{\Omega_i}{\sum_{m=1}^{M} \max \left\{ (|\lambda_1|)_m, (|\lambda_4|)_m \right\} \Gamma^m} &\text{,}\quad \Delta \tau^v_i \leq \dfrac{\Omega_i^2}{\frac{8}{3} \sum_{m=1}^{M}\frac{\mu_m}{\rho_m}(\Gamma^m)^2} \\
		\Delta \tau^g_i \leq \sqrt{\dfrac{h_i}{|\mathbf{g}|}} \quad\text{and}\quad \Delta \tau^s_i &\leq \sqrt{\dfrac{(\rho_1 + \rho_2)h^3_i}{4\pi \sigma}}
	\end{split}
\end{equation*}
where $h_i$ is the characteristic cell size. The eigenvalues $\lambda_1$ and $\lambda_4$ are computed as \eqref{eq:num_eigenvalues}. In the present work, the local characteristic cell size is taken as
\begin{equation}\label{eq:num_cellSize}
	h_i = K\dfrac{\Omega_i}{P_i} 
\end{equation}
where $P_i$ is the perimeter of the $i^\text{th}$ cell. The factor $K$ is taken as 4 for triangular cells and 3 for quadrilateral cells. The factors $K$ are chosen such that the local characteristic cell size is less than its average edge length. Thus, the interface thickness is minimized without compromising the stability of level set advection or reinitialization \cite{olsson2007conservative}. The global time-step is taken as
\begin{equation}\label{eq:num_timeStepGlobal}
	\Delta t = \text{CFL}\,\min_{i}\{\Delta t_i\} 
\end{equation}
where the Courant number $\text{CFL}$ is a simulation dependent parameter.

\section{Numerical results and discussions}\label{sec:res}

The efficacy of the proposed WC solver is tested on several two-dimensional problems. The incompressible two-phase problems are solved on structured as well as unstructured grids to demonstrate the adaptability of the solver. The unstructured meshes are generated using an open-source mesh generator Gmsh \cite{geuzaine2009gmsh}. The present study considers two grids equivalent if their average cell sizes \eqref{eq:num_cellSize} are similar as they result in similar interface thickness, dictated by the mesh dependent parameter \eqref{eq:num_epsilon}. For a domain discretized using two \emph{fairly} uniform meshes consisting entirely of $N_q$ quadrilateral cells and $N_t$ triangular cells respectively, it can be shown that the number of cells in the two meshes are related as $N_t \approx 1.4 N_q$, for the two meshes to have similar average cell sizes \eqref{eq:num_cellSize}. In the above approximation, the quadrilateral cells are assumed to be squares and the triangular cells to be equilateral triangles.

The problems are chosen such that the robustness of various constituent models of the proposed solver can be investigated. The results from the simulations are compared against analytical, experimental and numerical results reported in literature. In the qualitative results, the fluid interface is represented by the 0.5 contour of the level set function. In some test cases, additional interface contours at levels 0.05 and 0.95 is also plotted to demonstrate the effectiveness of the algorithm in maintaining the sharpness of the interface. The conservative property of the present solver is also tested by analysing the evolution of conservation error in all the problems. The area bounded by the 0.5 level set contour is taken as the measure of conservation in this study \cite{olsson2007conservative}. The relative area error $E^t_A$ is defined as
\[ E^t_A = \dfrac{A_{\psi=0.5}^t - A_{\psi=0.5}^0}{A_{\psi=0.5}^0} \]
where $A_{\psi=0.5}^t$ and $A_{\psi=0.5}^0$ are the area bounded by the 0.5 level set contour at time $t > 0$ and $t = 0$ respectively.

\subsection{Low amplitude sloshing}

The low amplitude sloshing problem \cite{yang2014upwind} models water sloshing under gravity in a stationary tank. The domain of the problem is a $L \times L$ square, where $L = 0.1\,\mathrm{m}$. At $t = 0$, the air-water interface is defined as $y(x) = L/2 + (L/20)\cos(\pi x/L)$ with water of density $1000\,\mathrm{kg/m}^3$ below the interface and air of density $1\,\mathrm{kg/m}^3$ above it. The air-water interface is allowed to slosh back and forth till $t = 2.5\,\text{s}$. Under inviscid flow assumption, the theoretical first mode of sloshing frequency of the free surface can be obtained analytically \cite{tadjbakhsh1960standing}. Therefore, diffusive fluxes and surface tension effects are neglected in this problem by setting the viscosities of the two fluids $\mu_1 = \mu_2 = 0$ and the surface tension coefficient $\sigma = 0$ respectively. Slip wall boundary conditions are imposed at all four walls. The acceleration due to gravity is taken as ${\mathbf{g}} = (0,-9.81)\,\mathrm{m/s^2}$. The initial velocity and pressure are set to zero in the entire domain. The sloshing problem is an apt test case to examine the fidelity of the proposed Riemann solver and the non-conservative discretization.

\begin{figure}[ht!]
	\centering
	\includegraphics[width=0.9\textwidth]{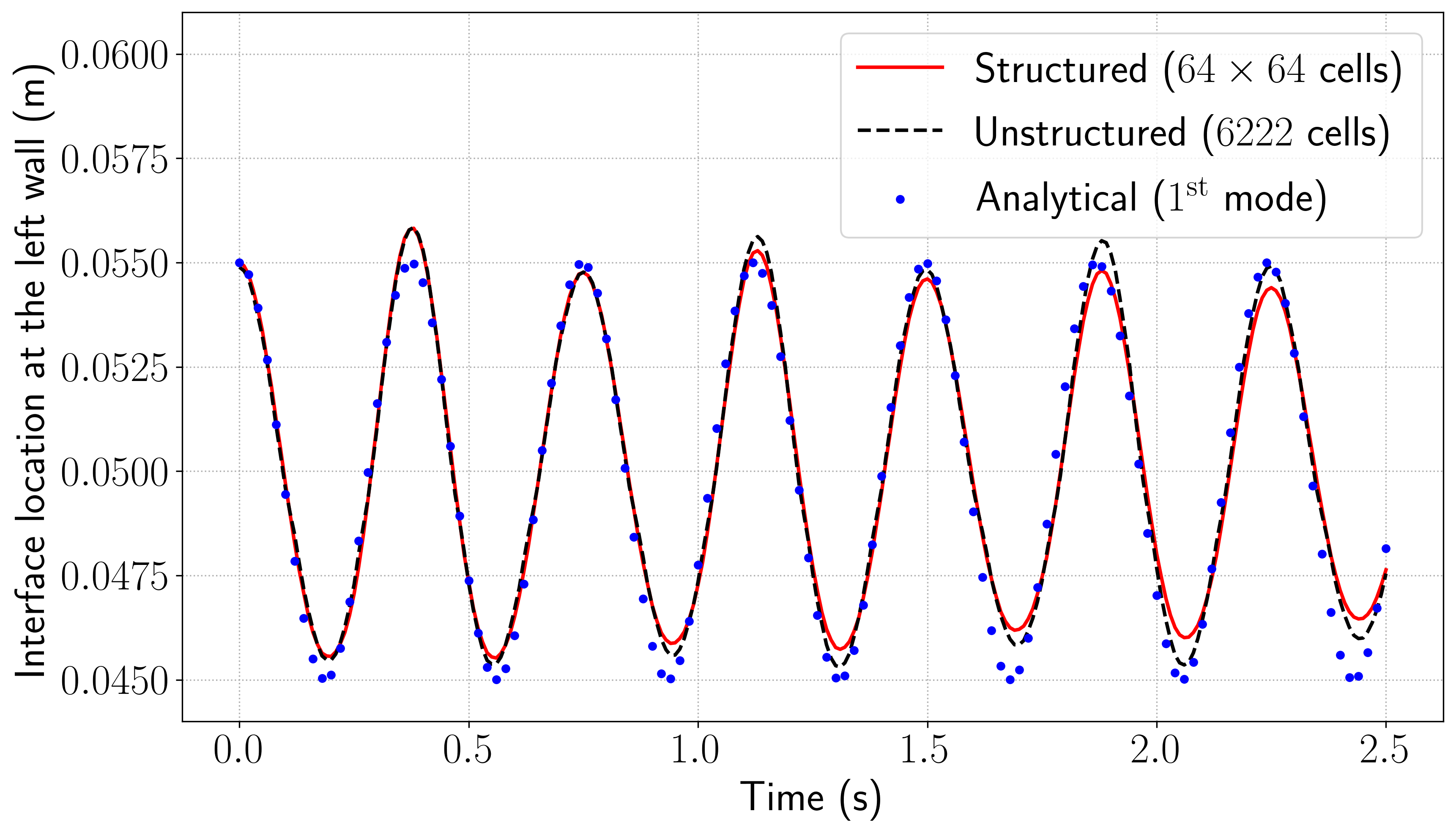}
	\caption{Location of air-water interface at the left wall compared against analytical first mode of oscillation.}
	\label{fig:res_slosh_freq}
\end{figure}

\begin{figure}[ht!]
	\centering
	\includegraphics[width=0.9\textwidth]{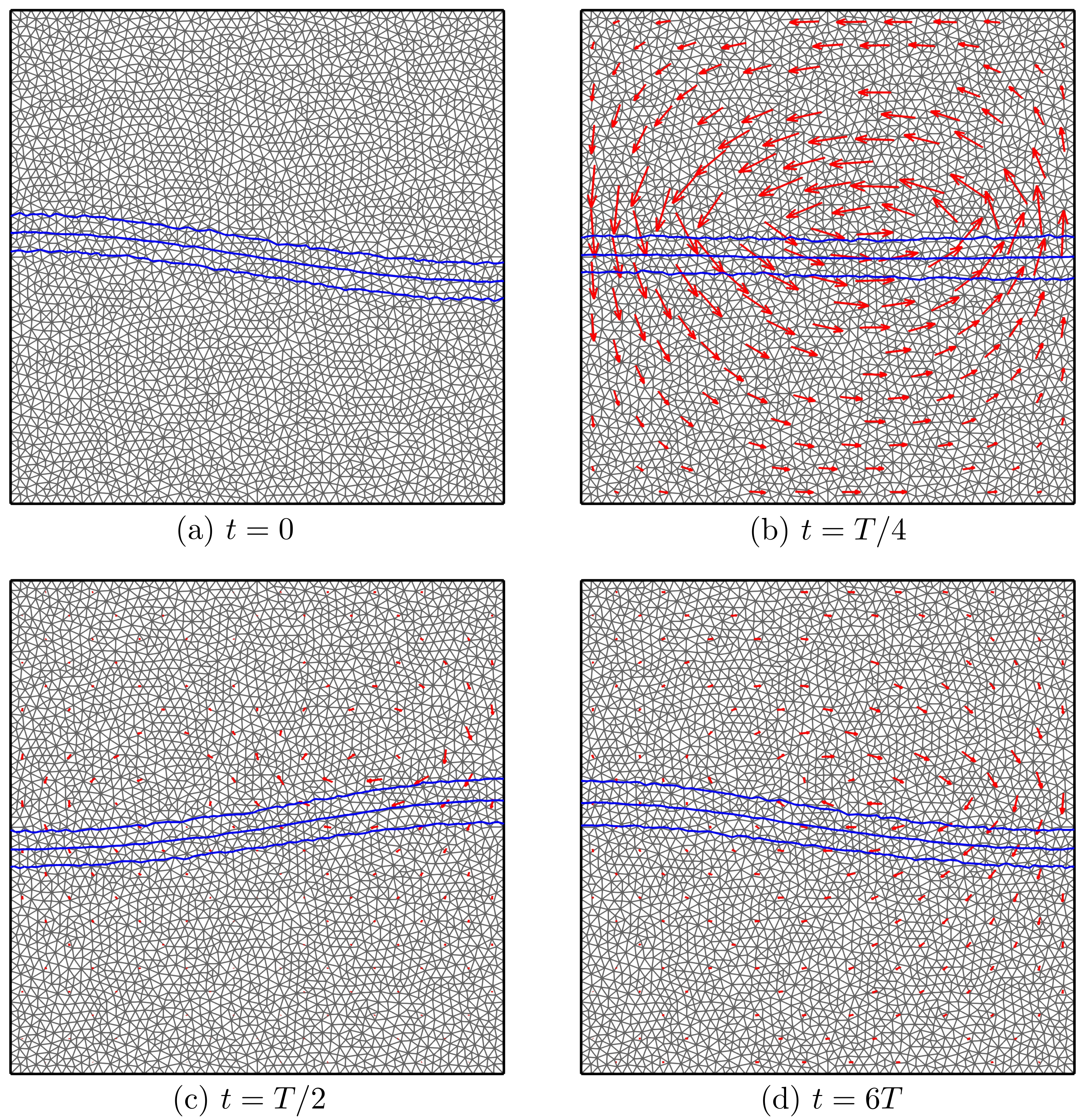}
	
	\caption{Evolution of air-water interface from the low amplitude sloshing simulation. Contours of level set function (blue) at $\psi = 0.05$, $0.5$ and $0.95$ are plotted. The represented velocity vectors (red) are scaled to $0.1$ times its magnitude.}
	\label{fig:res_slosh_contours}
\end{figure}

The numerical simulations are carried out on a $64 \times 64$ uniform Cartesian mesh as well as an equivalent unstructured mesh with $6222$ triangular cells. The AC parameter $\beta$ is taken as $1\times 10^3$ and CFL $= 0.9$. Four iterations of reinitialization were preformed with $C_\tau = 0.1$ after every 0.01 s. The analytical time period of oscillation is calculated as \cite{tadjbakhsh1960standing}
\begin{equation*}
	T = \dfrac{2 \pi}{\sqrt{|{\mathbf{g}}|k\tanh(kL/2)}}
\end{equation*}
where the wave number is defined as $k = \pi/L\,\text{m}^{-1}$. For the current test case, the time period is obtained as $ T \approx 0.373723\,\text{s}$.

The $y$-coordinate of the air-water interface at the left wall is plotted over time in figure \ref{fig:res_slosh_freq}. The results from the structured and unstructured grids agree well with the theoretical first mode oscillations. It should be noted that the alternate overshoots and undershoots with respect to the analytical first mode oscillations are due to the presence of second and higher modes of oscillation in the sloshing problem. Similar observations have also been reported in literature \cite{tadjbakhsh1960standing, yang2014upwind}. The location of the interface along with the instantaneous velocity vectors at four different times are shown in figure \ref{fig:res_slosh_contours}. For concision, the contours obtained only from the unstructured mesh is shown. The interface sharpness is preserved throughout the simulation demonstrating the efficacy of the reinitialization procedure. Furthermore, due to the low dissipative nature of the proposed HLLC Riemann solver coupled with the scheme used for the non-conservative terms, only a few reinitialization iterations are required, to preserve the interface thickness.

\begin{figure}[ht!]
	\centering
	\includegraphics[width=0.9\textwidth]{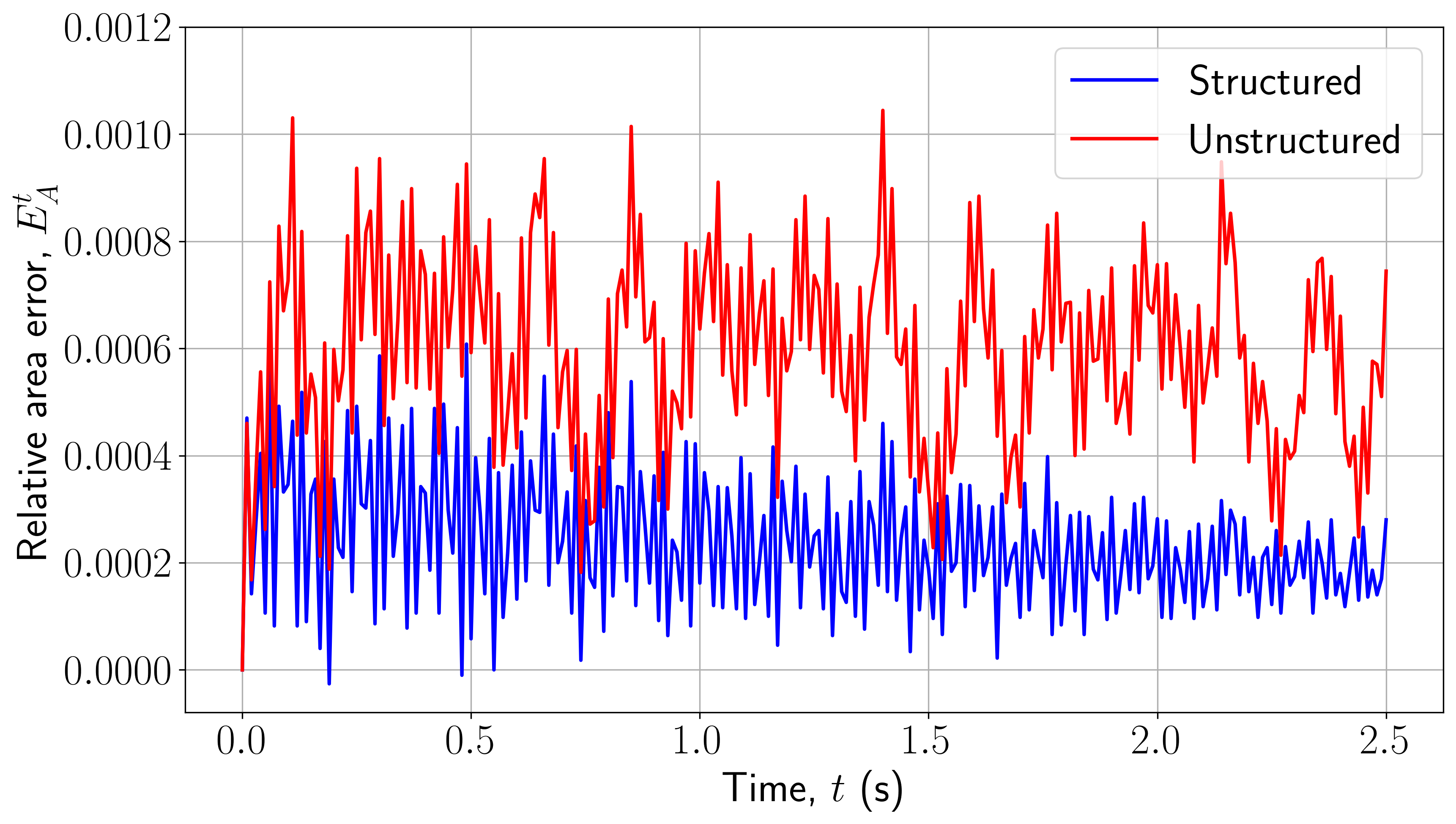}
	\caption{Relative conservation errors from low amplitude sloshing simulations.}
	\label{fig:res_slosh_area}
\end{figure}

The evolution of relative area error over time, from the simulations on structured and unstructured meshes is plotted in figure \ref{fig:res_slosh_area}, showing excellent area conservation.

\subsubsection{Grid convergence study on structured grids}

\begin{figure}[H]
	\centering
	\includegraphics[width=0.8\textwidth]{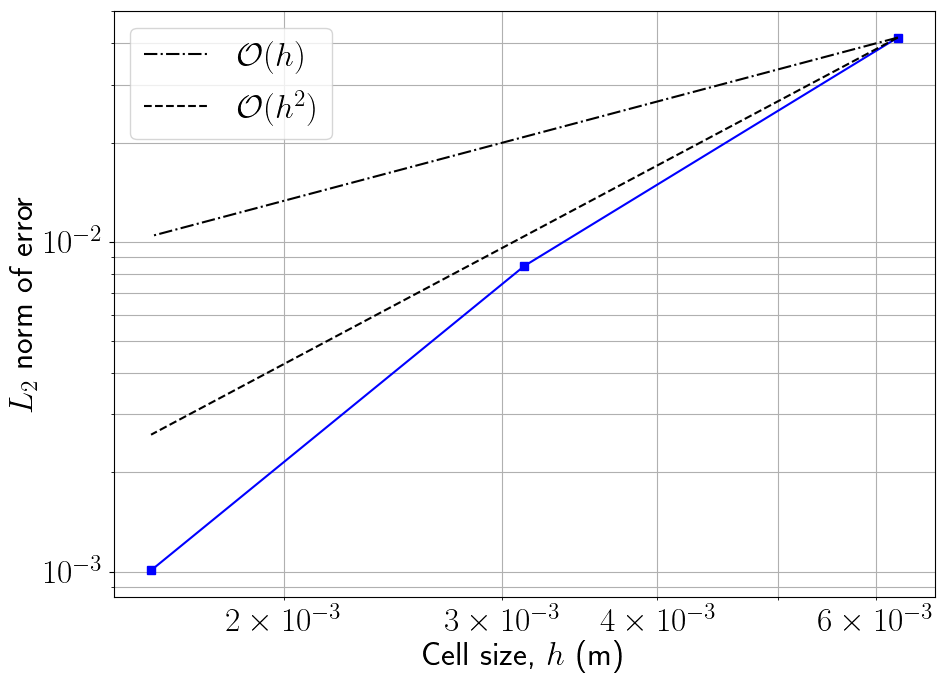}
	\caption{$L_2$ error convergence in low amplitude sloshing.}
	\label{fig:res_slosh_convergence_l2}
\end{figure}

To ensure that the spatial accuracy of the proposed WC solver is second order, a grid convergence study is performed. In this study, the low amplitude sloshing problem is simulated on uniform Cartesian grids of sizes: $16 \times 16$, $32 \times 32$, $64 \times 64$ and $128 \times 128$ cells. The data considered for grid convergence study is the $y$-coordinate of the fluid interface at the left wall over one time period $T$. For a given grid, the $L_1$, $L_2$ and $L_\infty$ norms of the error in data with respect to the fine mesh ($128 \times 128$ cells) result are computed and tabulated in table \ref{tab:res_slosh_convergence}. The $L_2$ norm of the error for different cell sizes $h$ is plotted in figure \ref{fig:res_slosh_convergence_l2}.

\begin{table}[H]
	\centering
	\begin{tabular}{| c | c | c | c | c |}
		\hline
		Error norm & $h = 1/16$ & $h = 1/32$ & $h = 1/64$ & Order \\
		\hline\hline
		$L_1$ & $0.6159038$ & $0.1219933$ & $0.0171164$ & $2.58$ \\ 
		\hline
		$L_2$ & $0.0416563$ & $0.0084593$ & $0.0010098$ & $2.68$ \\ 
		\hline
		$L_\infty$ & $0.0052188$ & $0.0012693$ & $0.0001122$ & $2.77$ \\
		\hline
	\end{tabular}
	\caption{Convergence order of norms of error in low amplitude sloshing.}
	\label{tab:res_slosh_convergence}
\end{table}

\subsection{Viscous dam break}

\begin{figure}[!htb]
	\centering
	\includegraphics[width=0.5\textwidth]{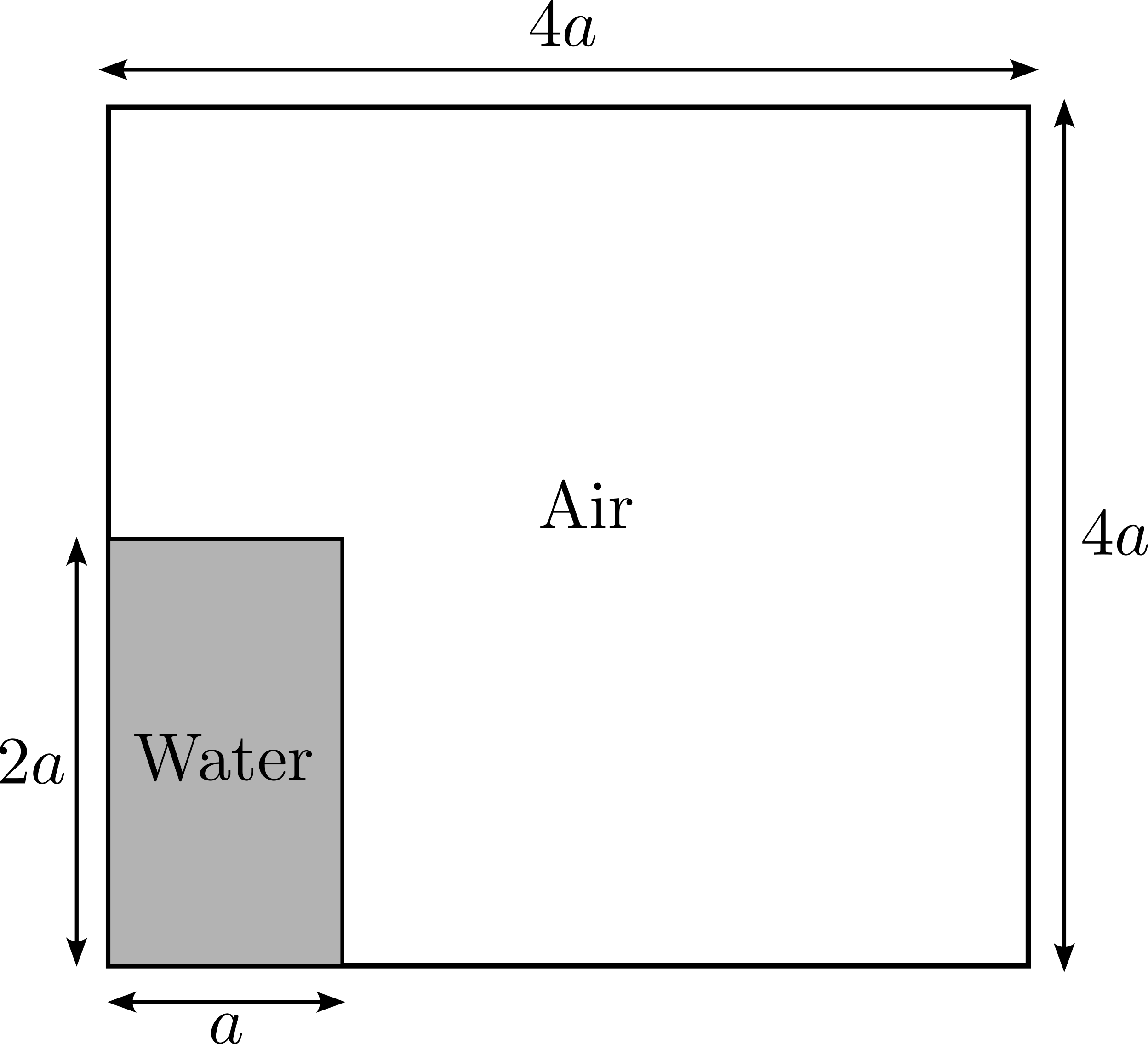}
	\caption{Schematic of the dam break problem ($a = 0.25$ m).}
	\label{fig:res_dambreak_schematic}
\end{figure}

The dam break problem is characterized by a high density ratio which leads to a violent flow rendering it a challenging test for any two-phase flow solver. Additionally, viscous effects are included in this test case to examine the efficacy of the scheme used to compute diffusive fluxes in the present work. In this problem \cite{zhang2010numerical}, a column of water rests on the left side of an enclosed $4a \times 4a$ m tank, where $a = 0.25$ m. The remaining volume of the tank is occupied by air. A schematic of the problem is shown in figure \ref{fig:res_dambreak_schematic}. At $t = 0$, the fictitious dam that holds the water column breaks and the water column is allowed to collapse under gravity. The flow is observed till time $t = 1.25$ s. During the collapse, the water surges along the bottom wall before hitting the right wall and then travels back. The initial column of water has a width $a$ and height $2a$. The density and dynamic viscosity of water is taken as $1000\,\mathrm{kg/m}^3$ and $1 \times 10^{-3} \,$kg/m-s respectively. The same for air is taken as $1\,\mathrm{kg/m}^3$ and $1.8 \times 10^{-5} \,$kg/m-s respectively. The acceleration due to gravity is taken as ${\mathbf{g}} = (0,-9.81)\,\mathrm{m/s^2}$. The domain is initialized with zero velocity while hydrostatic pressure based on gravity is set at $t = 0$. No-slip boundary conditions are imposed on all four walls.

The dam break simulations are carried out on a $80 \times 80$ uniform Cartesian mesh as well as an equivalent unstructured mesh with $8968$ triangular cells. The AC parameter $\beta = 1\times 10^4$ is considered and the Courant number is taken as CFL $= 0.8$. Owing to the violent nature of the flow, the air-water interface smears rapidly in this problem. Therefore, the level set field was reinitialized with five iteration after every $0.005$ s. A lower value of $C_\tau = 0.01$ was chosen to avoid any undesired interface movement due to frequent reinitialization \cite{parameswaran2023stable}.

\begin{figure}[H]
	\centering
	\includegraphics[width=0.9\textwidth]{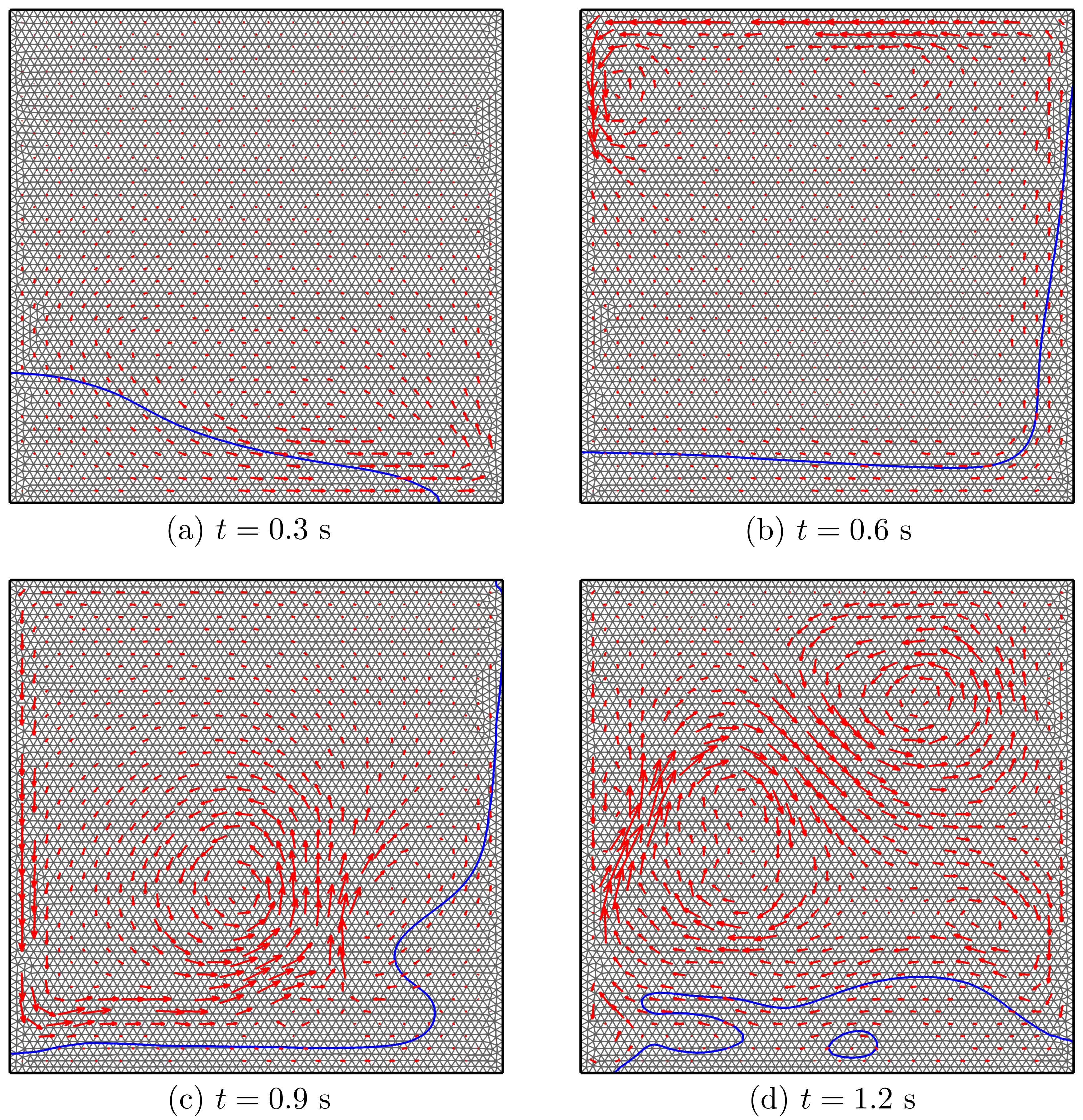}
	
	\caption{Evolution of air-water interface (blue), represented by contours of level set function at  $\psi = 0.5$, from the dam break simulation. The represented velocity vectors (red) are scaled $0.01$ times its magnitude.}
	\label{fig:res_dambreak_contours}
\end{figure}

\begin{figure}[H]
	\centering
	\includegraphics[width=\textwidth]{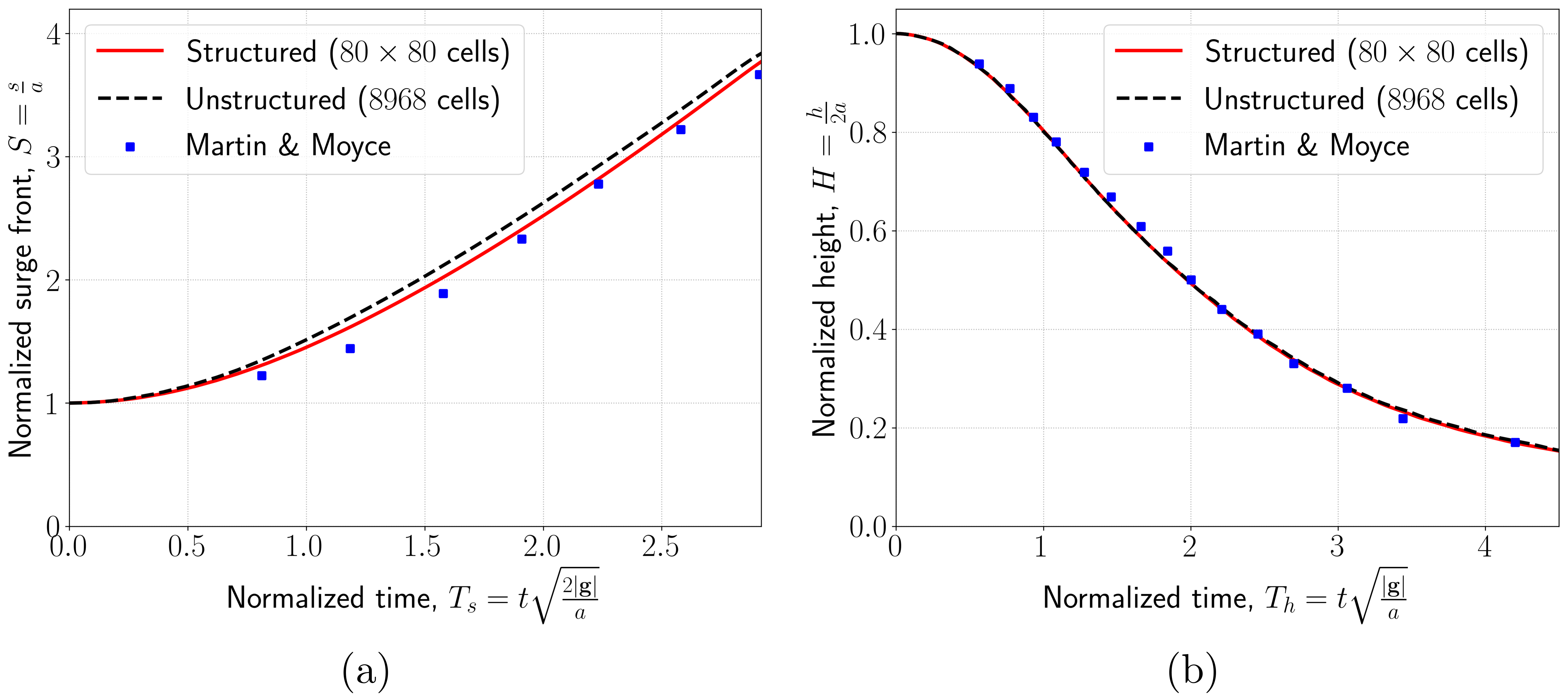}
	\caption{Evolution of the water column: (a) surge front along the bottom wall and (b) height at the left wall compared against experimental results \cite{martin1952part}.}
	\label{fig:res_damBreak_hs}
\end{figure}

The air-water interface, obtained from the simulation on the unstructured mesh, at four different times is shown in figure \ref{fig:res_dambreak_contours}. The contours obtained from structured mesh are not presented here for concision. Despite the general coarseness of the mesh used, the solver is capable of capturing intricate features of the dam break flow such as air pockets. The experimental data for the evolution of air-water interface along the right and bottom walls are available in literature \cite{martin1952part,koshizuka1996moving}. The position of the surge front at the bottom wall and height at the left wall as the water column collapses were plotted and compared with the experimental results \cite{martin1952part} in figure \ref{fig:res_damBreak_hs}. The results from the structured as well as unstructured case agrees well with the experimental results.

\begin{figure}[H]
	\centering
	\includegraphics[width=0.9\textwidth]{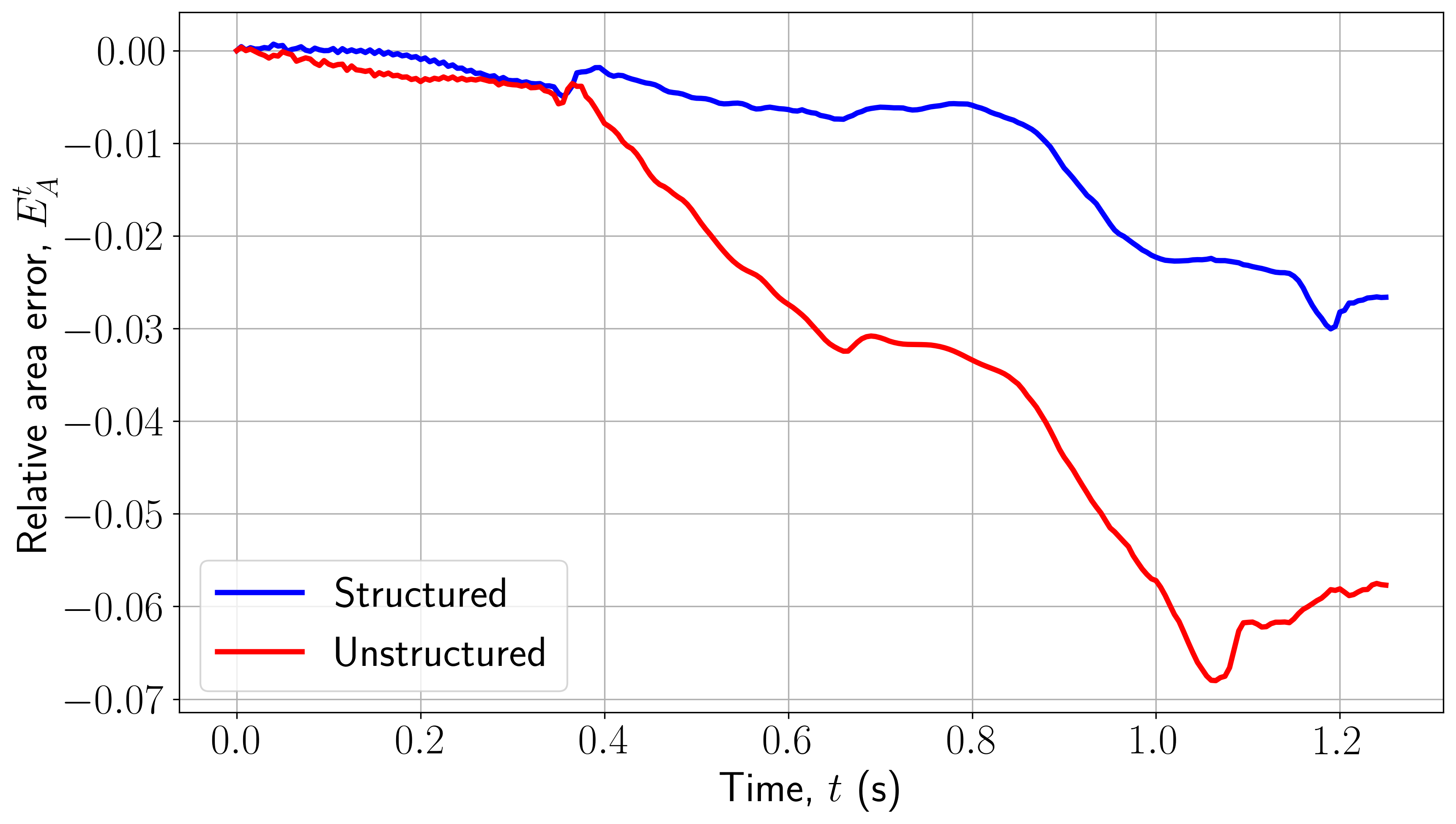}
	\caption{Relative conservation errors from dam break simulations.}
	\label{fig:res_dambreak_area}
\end{figure}

The evolution of relative area error over time, from the simulations on structured and unstructured meshes is plotted in figure \ref{fig:res_dambreak_area}. The plot shows good area conservation albeit higher relative errors compared to the ones from low amplitude sloshing problem.

\subsection{Rayleigh-Taylor instability}

\begin{figure}[ht!]
	\centering
	\includegraphics[width=0.3\textwidth]{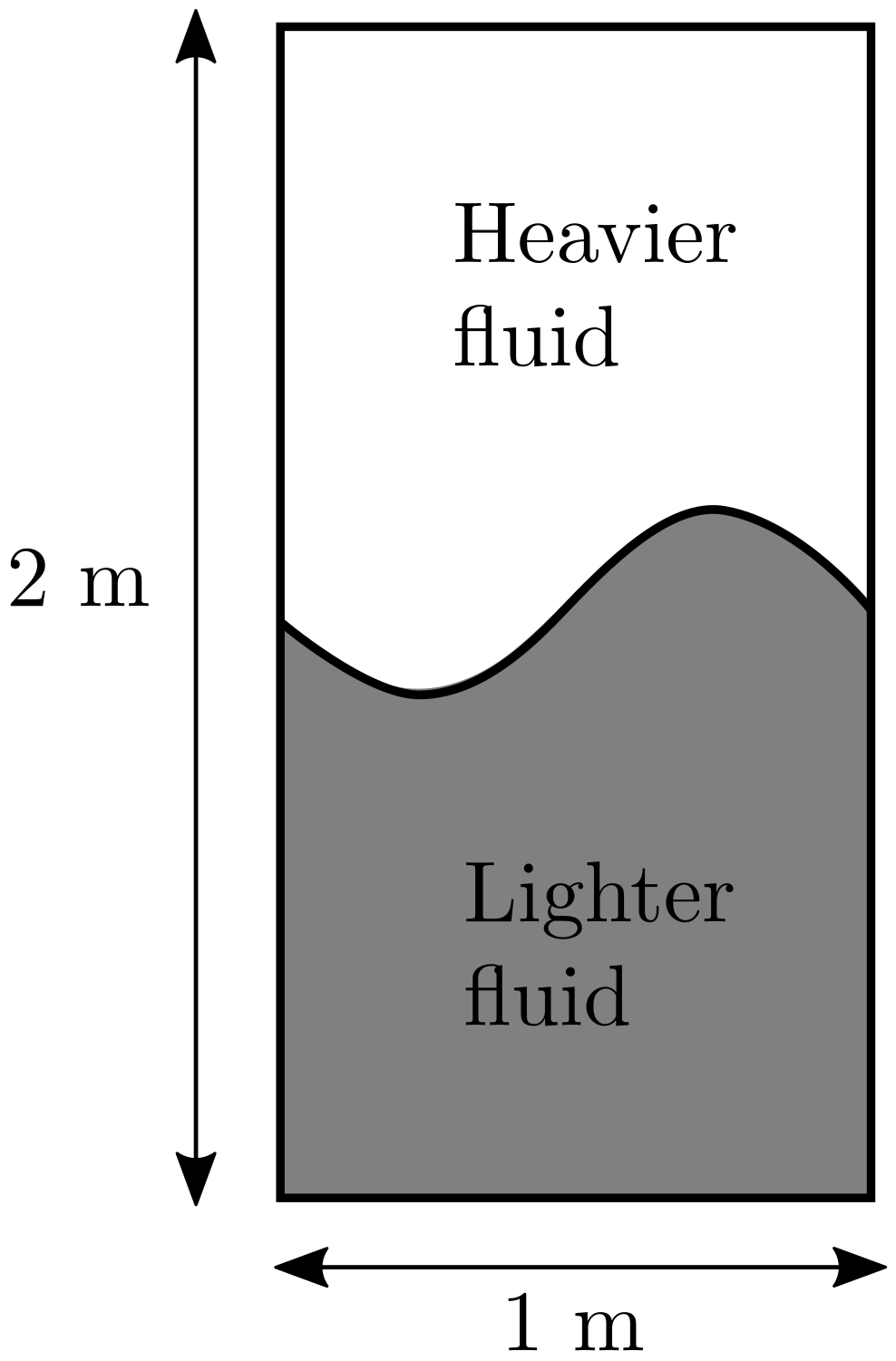}
	\caption{Schematic of the Rayleigh-Taylor instability problem.}
	\label{fig:res_rti_schematic}
\end{figure}

The Rayleigh-Taylor instability (RTI), which involves mixing of two immiscible fluids, poses a tremendous challenge to any interface-capturing method \cite{schilling2020progress}. The problem consists of a heavier fluid resting above a lighter fluid which, under gravity, forms an unstable system. The two-dimensional RTI problem proposed in \cite{cummins1999sph} is tested against the WC solver developed in the present work. Several benchmark numerical results for this particular case of RTI are available in literature \cite{pahar2016mixed,garoosi2021numerical,garoosi2022numerical}. The problem consists of a $H \times 2H$ rectangular domain, where $H = 1$ m, with a heavier fluid of density $\rho_1 = 1.8\,\mathrm{kg/m}^3$ resting above a lighter fluid of density $\rho_2 = 1\,\mathrm{kg/m}^3$. The dynamic viscosities of the heavier and lighter fluid are $\mu_1 = 1.8 \times 10^{-2} \,$kg/m-s and $\mu_2 = 1 \times 10^{-2} \,$kg/m-s respectively. The nominal gravitational acceleration of ${\mathbf{g}} = (0,-17.64)\,\mathrm{m/s^2}$ is considered \cite{garoosi2022numerical}, such that the fluid flow is governed by Atwood number $At = (\rho_1 - \rho_2)/(\rho_1 + \rho_2) = 2/7$ and Reynolds number $Re = \rho_1H\sqrt{|{\mathbf{g}}|H}/\mu_1 = 420$ \cite{tryggvason2011direct}. At $t = 0$, the two fluids are separated by a sinusoidal interface specified as $y(x) = 1.0 - 0.15\sin (2\pi x)$ as shown in figure \ref{fig:res_rti_schematic}. No-slip boundary conditions are imposed on all four walls.

As time elapses, the heavier fluid moves down while the lighter fluid moves upward due to buoyancy, forming two counter-rotating vortices in the domain. Due to the small density differences along with the differences in tangential velocity across the interface, Kelvin-Helmholtz instability \cite{chandrasekhar2013hydrodynamic} develops and the interface becomes unstable forming a mushroom-like profile. The mushroom-like structure undergoes further transformation as the intensity of the two vortices increase.

The RTI problem is solved on a $128 \times 256$ uniform Cartesian grid. The domain is initialized with zero velocity while hydrostatic pressure based on gravity is set at $t = 0$. The AC parameter is taken as $\beta = 1\times 10^3$ and the Courant number is taken as CFL $= 0.9$. The simulations are carried out till $t = 1.2$ s and the level set field is reinitialized using $25$ iterations, with the stability parameter $C_\tau = 0.01$, after every every $t = 0.01$ s. Unlike the previous cases, maintaining the interface thickness throughout the simulation is paramount in order to capture all the intricate features of the RTI. Therefore, a relatively higher number of iterations per reinitialization step is employed in this simulation.

\begin{figure}[H]
	\centering
	\includegraphics[width=0.7\textwidth]{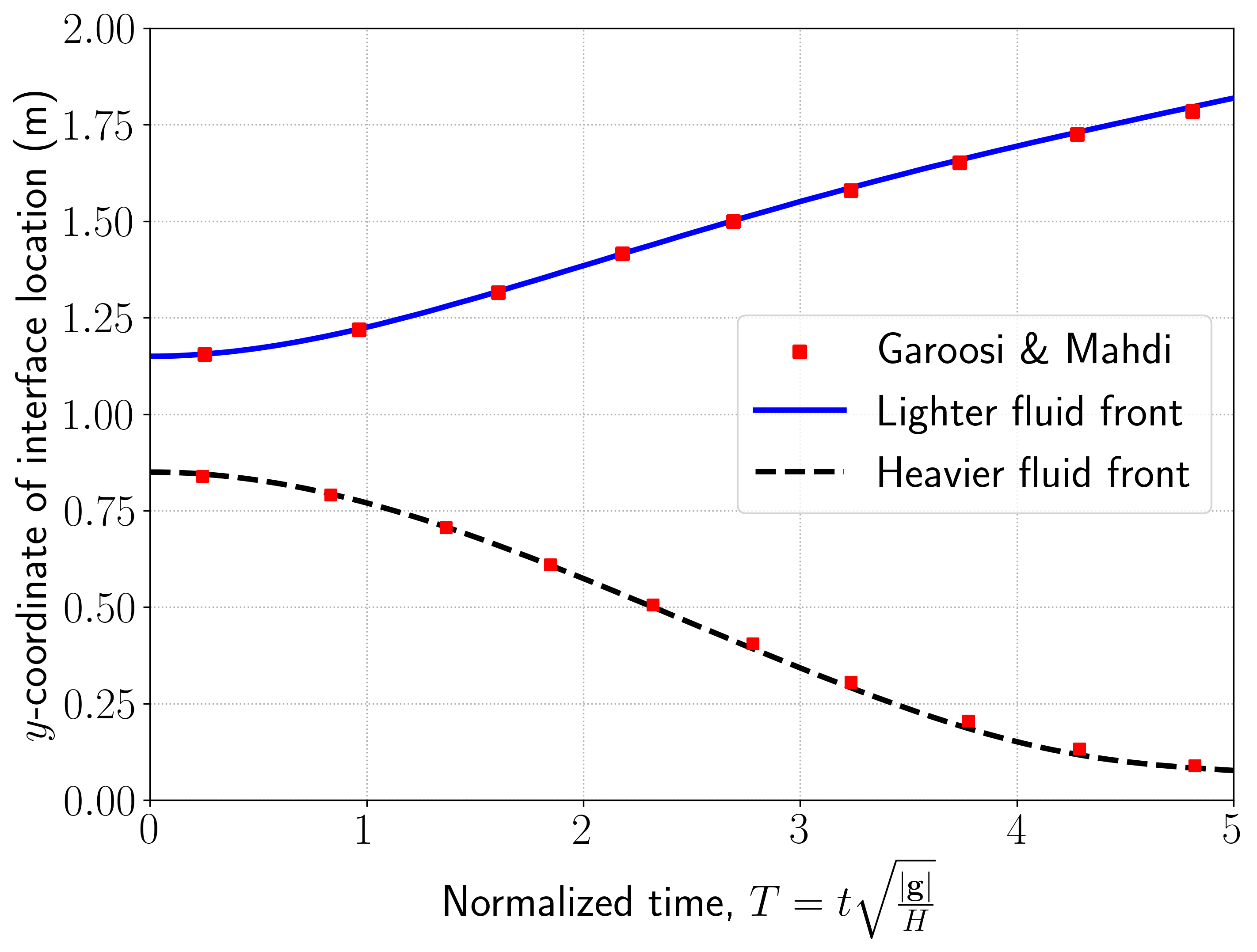}
	\caption{Evolution of the lighter and heavier fluid fronts compared against benchmark numerical results \cite{garoosi2022numerical}.}
	\label{fig:res_rti_front}
\end{figure}

\begin{figure}[ht!]
	\centering
	\includegraphics[width=0.9\textwidth]{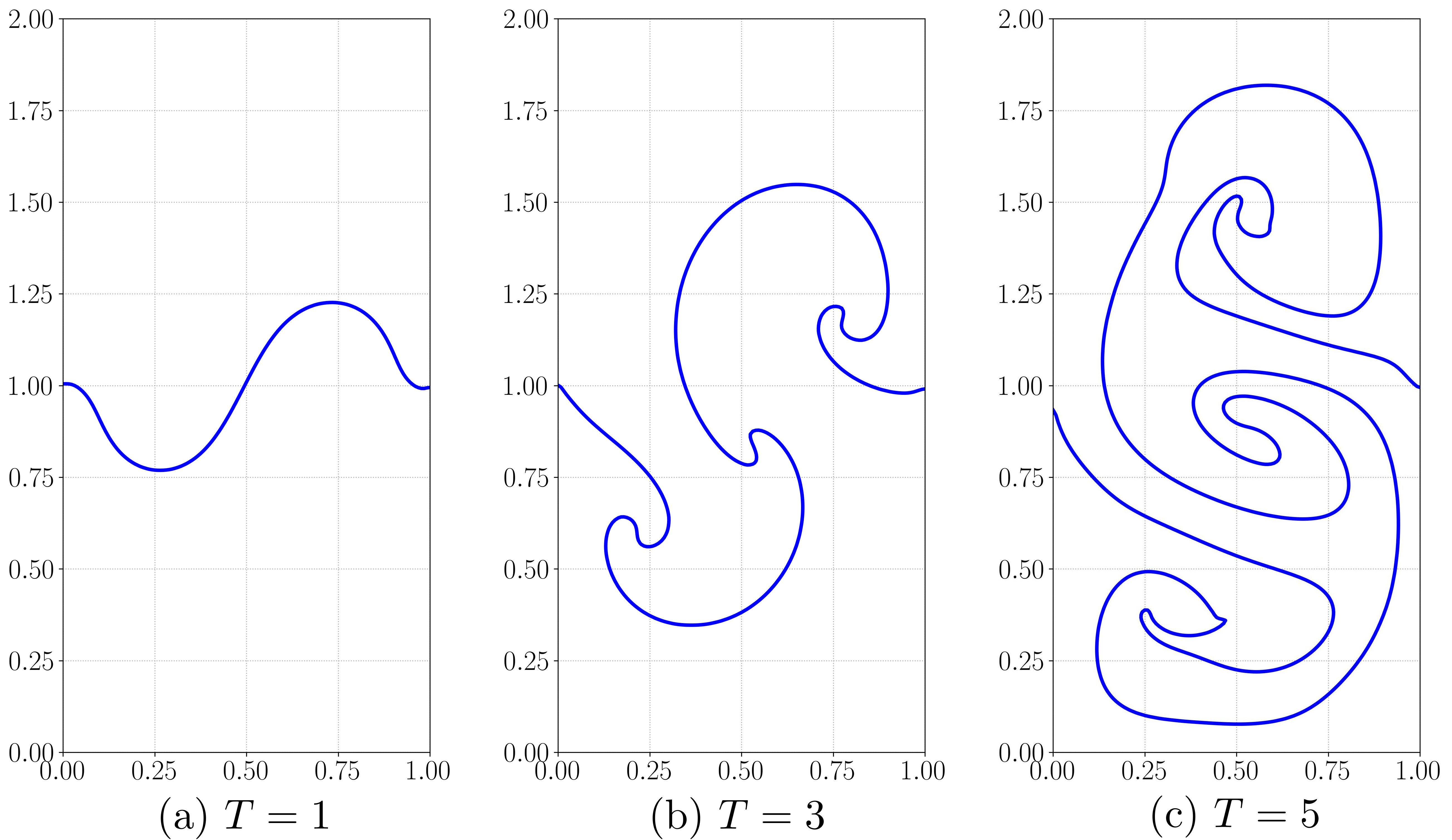}
	\caption{Evolution of the interface in the Rayleigh-Taylor instability. Here non-dimensional time $T = t\sqrt{|\mathbf{g}|/H}$.}
	\label{fig:res_rti_contours}
\end{figure}

\begin{figure}[ht!]
	\centering
	\includegraphics[width=0.9\textwidth]{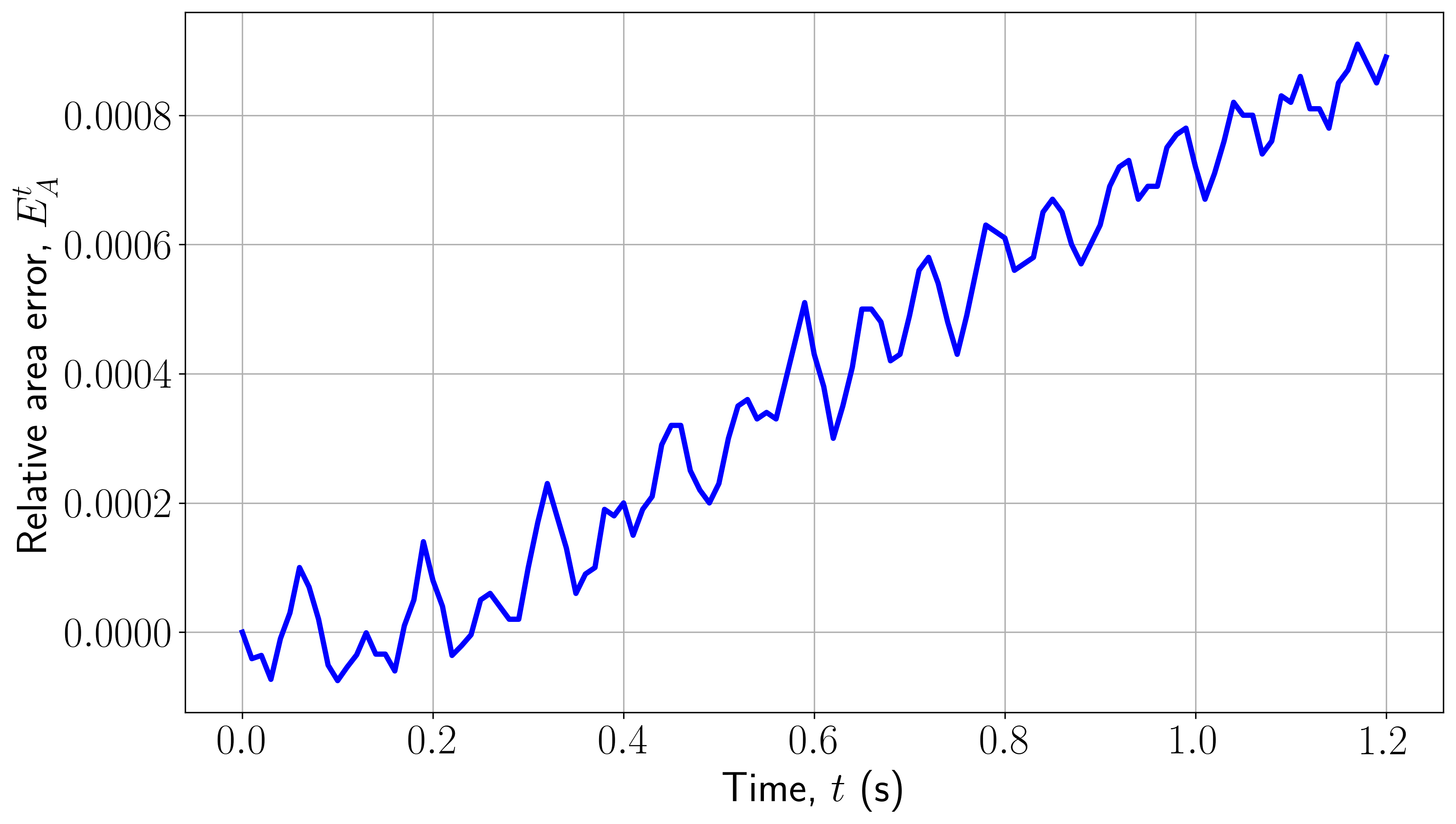}
	\caption{Relative conservation errors from Rayleigh-Taylor instability simulation.}
	\label{fig:res_rti_area}
\end{figure}

The location of the maximum and minimum positions ($y$-coordinate) of the interface is plotted over time in figure \ref{fig:res_rti_front}. The time history of lighter fluid front (maximum $y$-coordinate) and heavier fluid front (minimum $y$-coordinate) are in excellent agreement with literature \cite{pahar2016mixed,garoosi2021numerical,garoosi2022numerical}. The evolution of the fluid interface, at three different instances, is shown in figure \ref{fig:res_rti_contours}. Benchmark interface profiles obtained through mesh-free (particle) methods with $250 \times 500$ particles are provided in \cite{garoosi2021numerical}. The interface shapes obtained from a relatively coarser mesh using the present WC solver is almost indistinguishable from the benchmark results \cite{garoosi2021numerical}.

The evolution of relative area error over time is plotted in figure \ref{fig:res_rti_area}, showing excellent area conservation.

\subsection{Bubble rise in a column}\label{sec:bubblerise}

\begin{figure}[ht!]
	\centering
	\includegraphics[width=0.35\textwidth]{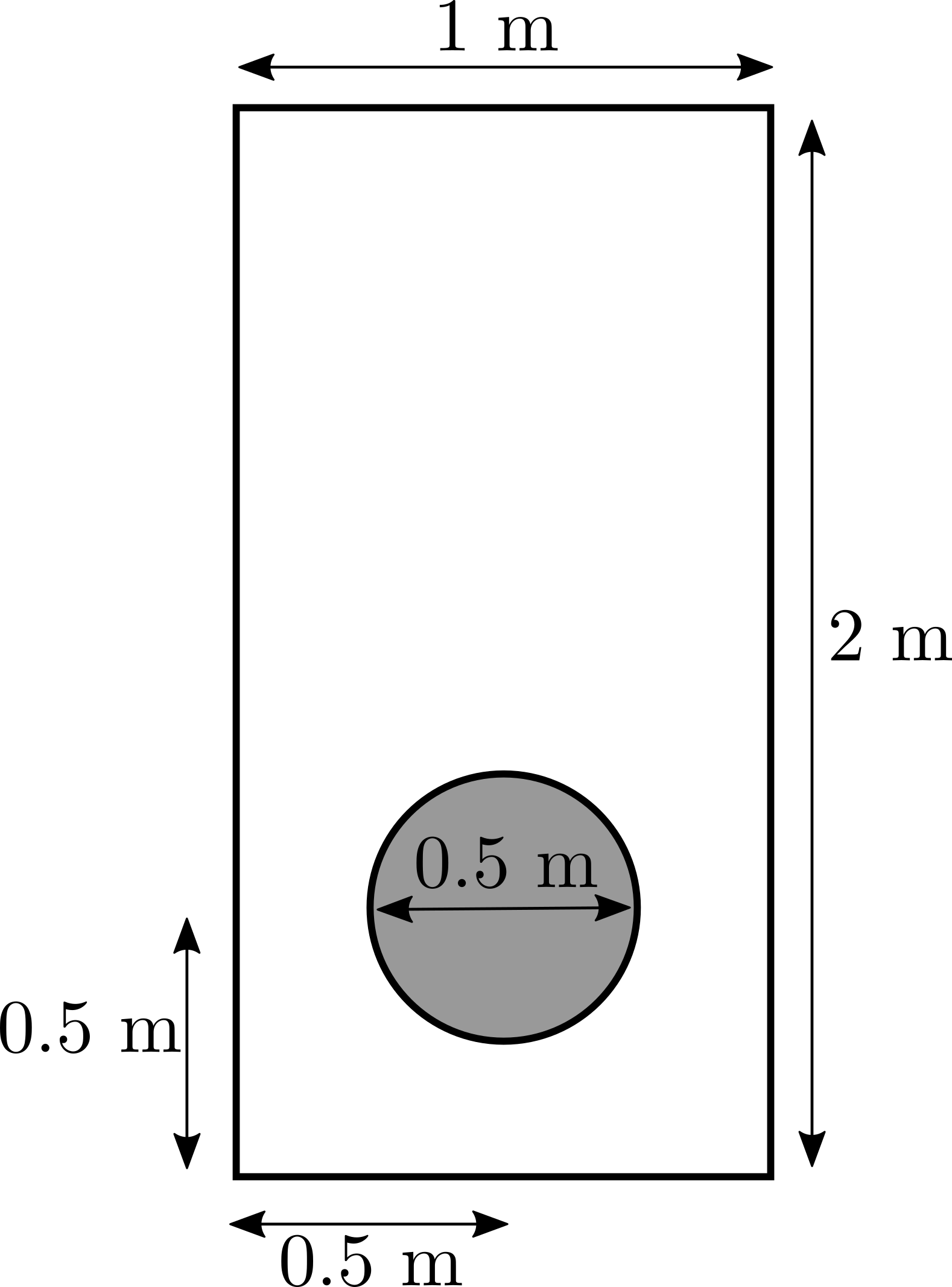}
	\caption{Schematic of the bubble rise in a column problem.}
	\label{fig:res_bubblerise_schematic}
\end{figure}

Consideration of surface tension is vital in any two-phase flow solver when applied to real-life engineering problems. Against this background, the problem of two-dimensional bubble rise in a column, proposed by Hysing et. al \cite{hysing2009quantitative} is considered in the present work. The problem consists of a circular bubble of diameter $d_0 = 0.5$ m centered at $(0.5,0.5)$ m inside a $1 \times 2$ m column filled with a heavier fluid. The schematic detailing the initial setup of the problem is shown in figure \ref{fig:res_bubblerise_schematic}. Due to buoyancy, the bubble rises up and its initial circular shape is deformed. The characteristic length $L$ is the same as the bubble diameter $d_0$ and the velocity scale is defined as $U = \sqrt{|{\mathbf{g}}|d_0}$, in this problem. The flow is governed by Reynolds number $Re$ and E\"{o}tv\"{o}s number $Eo$ defined as
\begin{equation*}
	Re = \dfrac{\rho_1 U L}{\mu_1} \quad\text{and}\quad Eo = \dfrac{\rho_1 U^2 L}{\sigma} 
\end{equation*}
where $(\cdot)_1$ are the material properties of the heavier fluid.

\begin{figure}[ht!]
	\centering
	\includegraphics[width=0.8\textwidth]{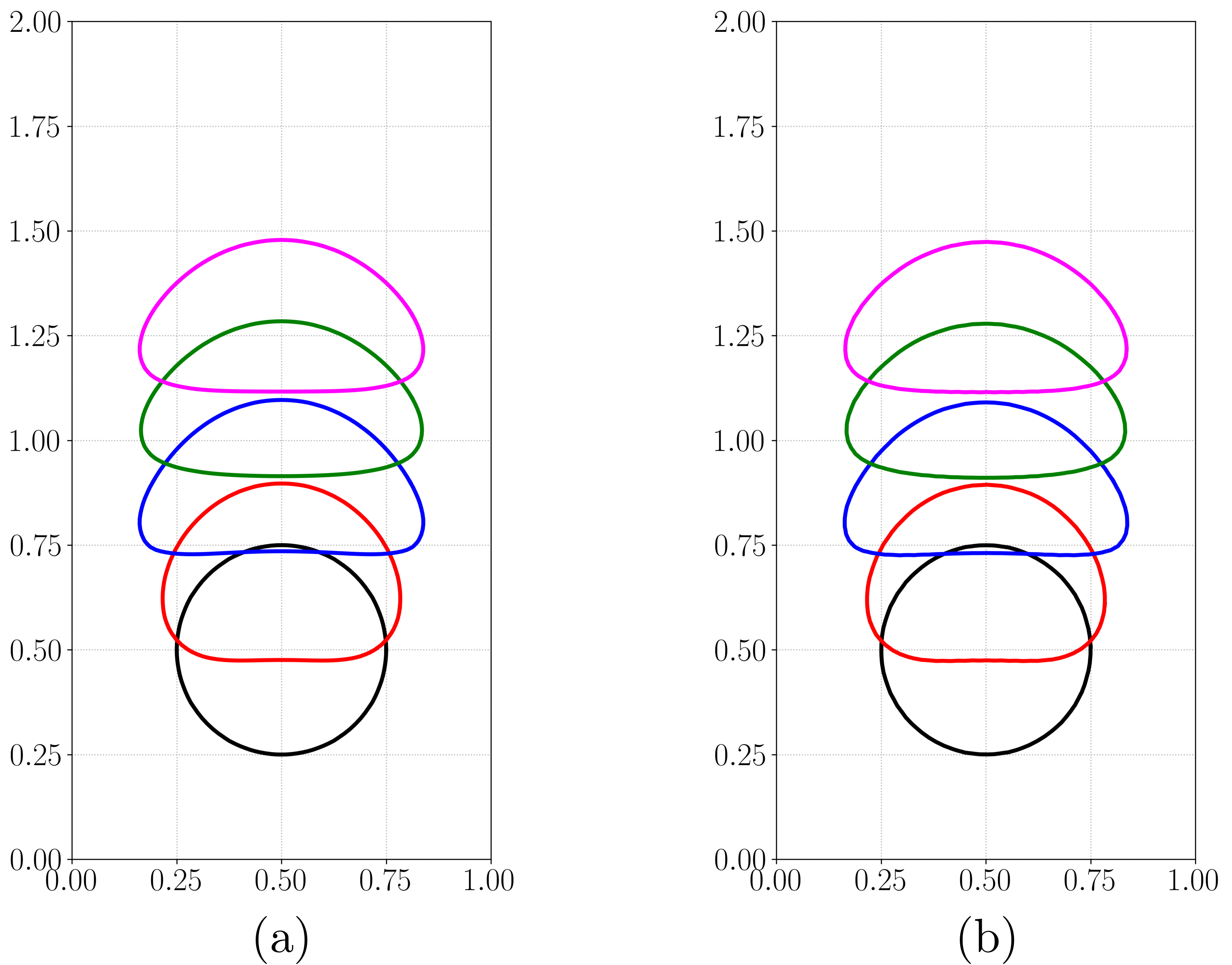}
	\caption{Evolution of the bubble interface, as it rises up in the column, obtained from simulations (case 1) on: (a) structured and (b) unstructured mesh. The contours of level set function $\psi = 0.5$ shown at $t = $ 0, 1, 2, 3 and 4 s.}
	\label{fig:res_bubblerise1_contours}
\end{figure}

\begin{figure}[ht!]
	\centering
	\includegraphics[width=0.6\textwidth]{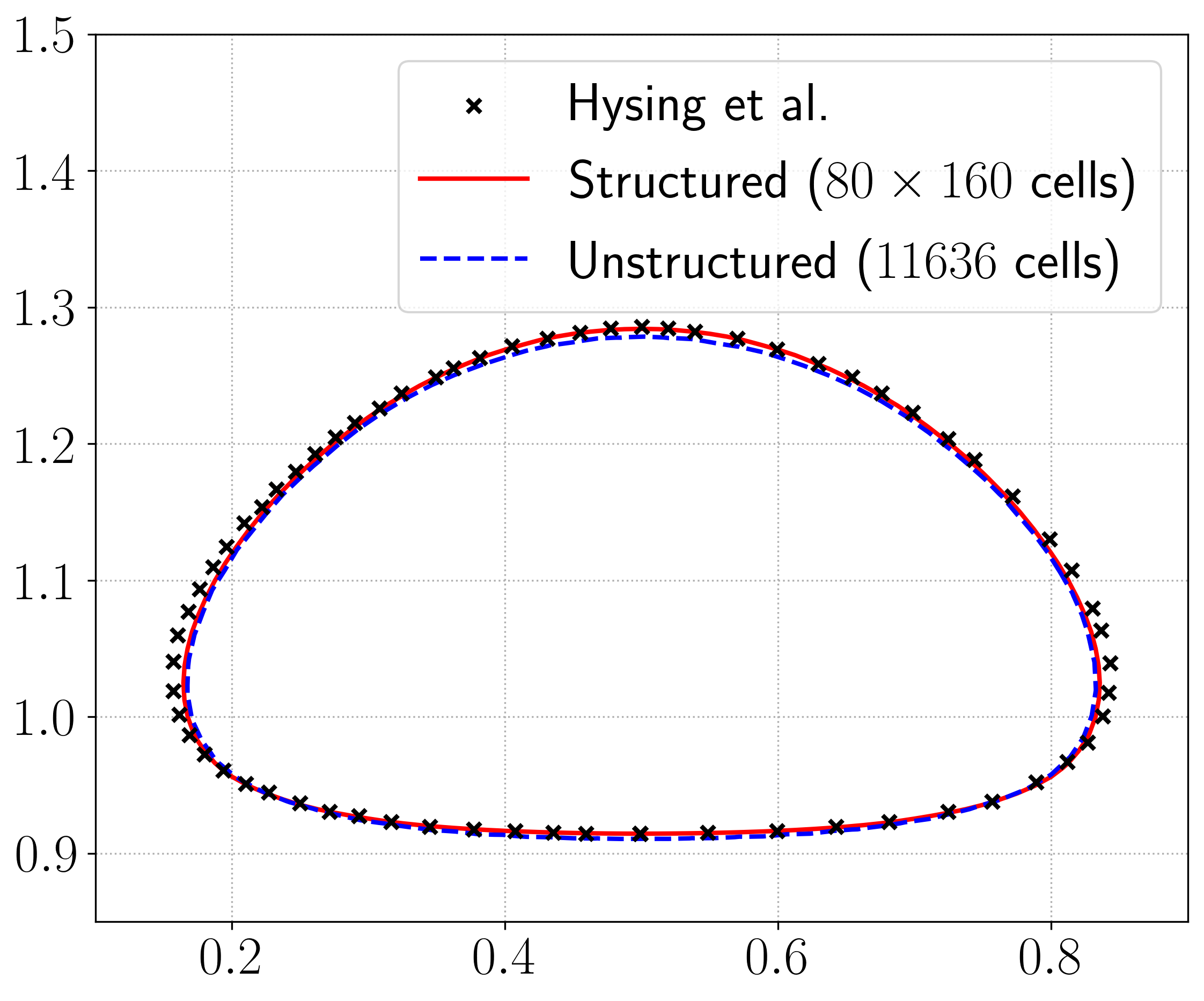}
	\caption{Bubble shape at time $t = 3$ s compared against benchmark numerical result \cite{hysing2009quantitative}.}
	\label{fig:res_bubblerise1_t_3}
\end{figure}

The densities of heavier and lighter fluids are taken as $\rho_1 = 1000\,\mathrm{kg/m}^3$ and $\rho_2 = 100\,\mathrm{kg/m}^3$ respectively. The dynamic viscosities of the heavier fluid is taken as $\mu_1 = 10$ kg/m-s and that of lighter fluid is taken as  $\mu_2 = 1$ kg/m-s.  The surface tension coefficient is 24.5 N/m. Therefore, the flow is governed by $Re = 35$ and $Eo = 10$ with density and viscosity ratios of 10. The bubble falls under the ellipsoid regime \cite{clift2005bubbles}, i.e., the surface tension effects are strong enough to hold the bubble together without resulting in any break up.

The domain is initialized with zero velocity while hydrostatic pressure based on gravity is set at $t = 0$. No-slip boundary conditions are imposed on the top and bottom walls while the left and right walls are considered to be slip walls. The simulations are carried out till $t = 4$ s with the Courant number $\text{CFL} = 0.8$. The level set field is reinitialized after every $0.01$ s though 20 iterations of reinitialization with stability parameter $C_\tau = 0.01$. The simulations are carried out an $80 \times 160$ uniform Cartesian grid as well as an unstructured grid with $11636$ cells. In this case, the two meshes are not equivalent as per the relation outlined earlier. The AC parameter is taken as $\beta = 1 \times 10^{3}$.

Results from the simulations are validated against the benchmark numerical results provided in \cite{hysing2009quantitative}. Along with the shape of bubble interface, represented by the contour of level set function $\psi = 0.5$, three other benchmark quantities as in \cite{hysing2009quantitative} are used to quantify the temporal evolution of the bubble. The computation of these three benchmark quantities, namely circularity, centroid location, and rise velocity, are detailed in \cite{parameswaran2023conservative}.

The evolution of the bubble interface as it rises up the column is shown in figure \ref{fig:res_bubblerise1_contours}. Almost indistinguishable bubble interface profiles are obtained from structured and unstructured grids. The bubble shapes obtained from the two simulations at $t = 3$ s, i.e., once the bubble attains terminal velocity, is compared against the benchmark shape obtained from fine mesh simulation \cite{hysing2009quantitative} in figure \ref{fig:res_bubblerise1_t_3}. The bubble shapes obtained on structured and unstructured mesh using the proposed solver agree well with the benchmark result.

\begin{figure}[H]
	\centering
	\includegraphics[width=\textwidth]{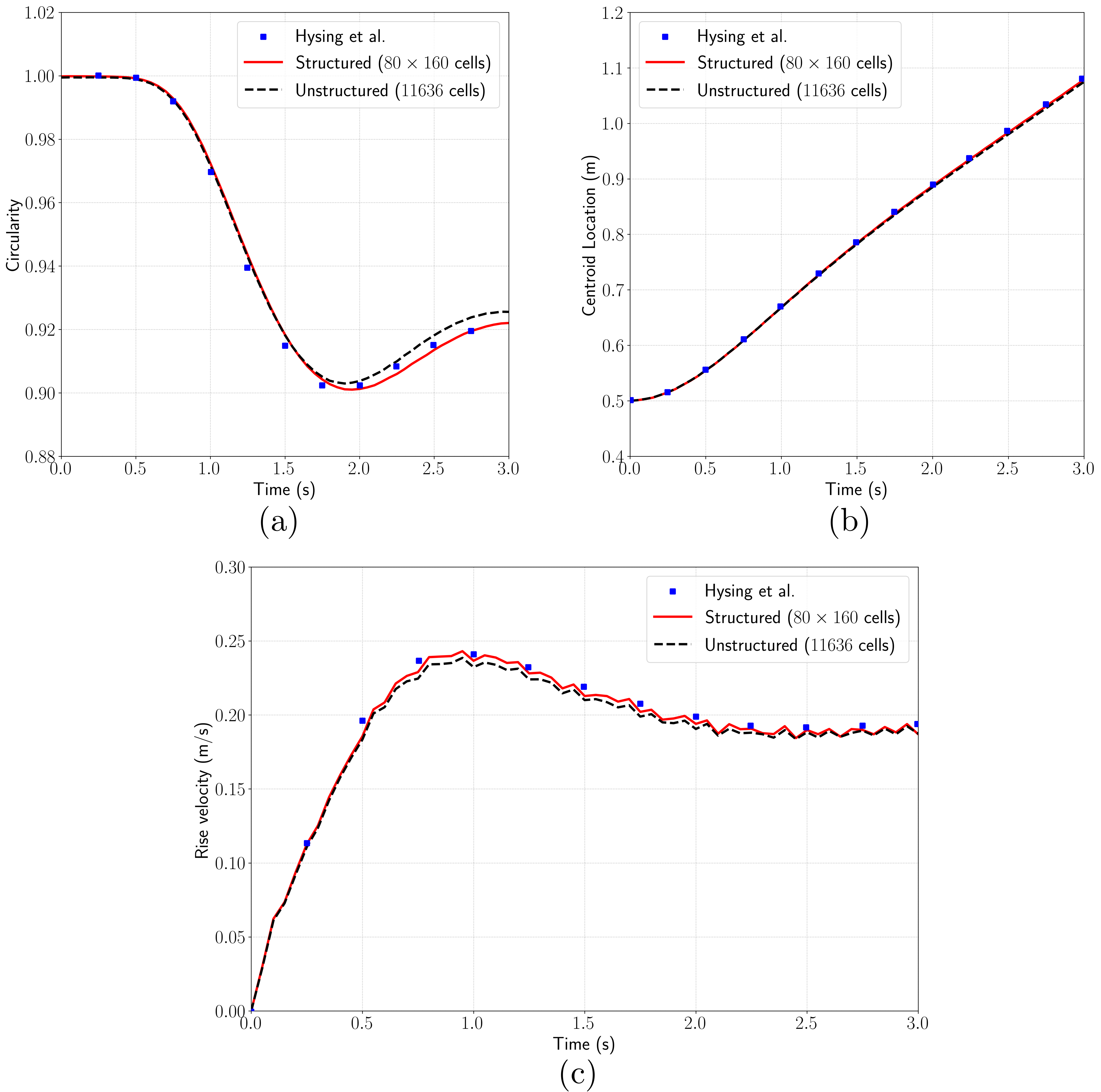}
	\caption{Evolution of benchmark quantities: (a) circularity, (b) centroid location, and (c) rise velocity for bubble rise in a column compared against numerical results in literature \cite{hysing2009quantitative}.}
	\label{fig:res_bubblerise1_plots}
\end{figure}

As mentioned earlier, along with the bubble shapes, the evolution of centroid location, rise velocity and circularity obtained from the simulations are also examined. Figure \ref{fig:res_bubblerise1_plots} compares these quantities with the benchmark results in \cite{hysing2009quantitative}. The evolution of the parameters follow the same trend as the benchmark results. The oscillations in the evolution of rise velocity can be attributed the pressure and velocity oscillations inherent in weakly compressible methods. The pressure and velocity oscillations in weakly compressible methods, well documented in literature \cite{inamuro2016improved,sitompul2019filtered}, are induced by the presence of acoustic waves in the system. Methods such as the evolving pressure projection method \cite{yang2021weakly,yang2022momentum} have been proposed in literature that can damp the acoustic waves and in turn the oscillations. However, additional iterations between physical time-steps are involved in these methods and is outside the scope of the present work.

\begin{figure}[ht!]
	\centering
	\includegraphics[width=0.9\textwidth]{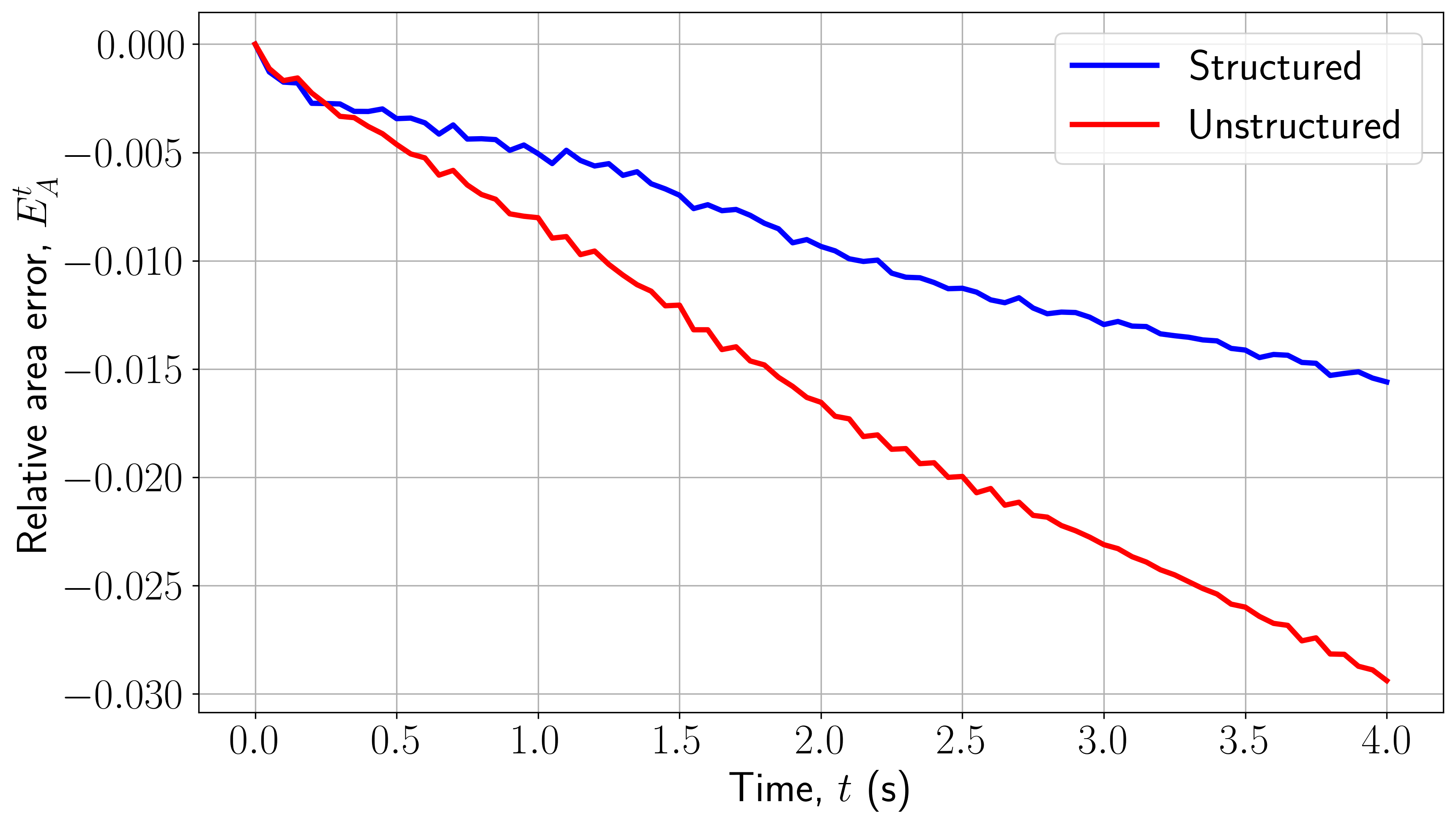}
	\caption{Relative conservation errors from bubble rise simulations.}
	\label{fig:res_bubblerise1_area}
\end{figure}

The evolution of relative area error over time, from the simulations on structured and unstructured meshes is plotted in figure \ref{fig:res_bubblerise1_area}, showing good area conservation.

\section{Conclusion}\label{sec:conc}

A fully-explicit weakly compressible (WC) solver to simulate incompressible two-phase flows has been developed in this work. The WC model couples the general pressure equation, momentum conservation equations and conservative level set advection equation. The WC model is shown to be hyperbolic in time and therefore admits the development of Riemann solvers for the system. A novel contact-preserving HLLC-type Riemann solver is formulated to compute the convective fluxes in the present solver. A simple novel oscillation-free scheme is derived to discretize the non-conservative terms present in the WC model. The discretization is dependent on the convective flux formulation subject to a steady-state constraint \cite{abgrall1996prevent}. 

The efficacy of the proposed solver is tested on a variety of incompressible two-phase flow problems. The finite volume solver is formulated for an arbitrary two-dimensional mesh and can therefore be implemented on a structured, unstructured or hybrid meshes. To demonstrate the adaptability of the present solver, it is tested on structured as well as unstructured meshes. The results from the simulations were compared against analytical, experimental and numerical results available in the literature. The results from inviscid low amplitude sloshing problem validated the accuracy of the novel HLLC Riemann solver and the proposed discretization of the non-conservative terms. The solver was further extended to include viscous effects and was evaluated using dam break and Rayleigh-Taylor instability problems. These simulations demonstrated the solver's robustness in handling violent flow phenomena and accurately resolving interfaces during the mixing of two immiscible fluids. Additionally, the solver's capability to incorporate surface tension effects was verified through the simulations of rising bubble in a column. The area conservation errors across all the test cases remained within acceptable limits.

\newpage


\printbibliography[title=References]

\end{document}